\documentclass[showpacs,pra,preprint,amsmath,amssymb,aps]{revtex4-1}
\usepackage{graphicx}
\usepackage{bm}

\begin{document}


\title{How tetraquarks can generate a second chiral phase transition}

\author{Robert D. Pisarski}
\email{pisarski@bnl.gov}
\affiliation{Department of Physics, Brookhaven National Laboratory,
Upton, NY 11973}
\affiliation{RIKEN/BNL, Brookhaven National Laboratory,
Upton, NY 11973}
\author{Vladimir V. Skokov}%
\email{vskokov@bnl.gov}
\affiliation{RIKEN/BNL, Brookhaven National Laboratory,
Upton, NY 11973}

\begin{abstract}
We consider how tetraquarks can affect the chiral phase transition
in theories like QCD, with light quarks coupled to three colors.
For two flavors the tetraquark
field is an isosinglet, and its effect is minimal.
For three flavors, however, the tetraquark field transforms in the same
representation of the chiral symmetry group
as the usual chiral order parameter,
and so for very light
quarks there {\it may} be two chiral phase transitions, 
which are both of first order.
In QCD, results from the lattice indicate that any transition
from the tetraquark condensate is a smooth crossover.
In the plane of temperature and quark chemical potential, though,
a crossover line for the tetraquark condensate 
is naturally related to the transition line
for color superconductivity.  
For four flavors we suggest that a triquark field, 
antisymmetric in both flavor and color, combine to form hexaquarks.
\end{abstract}
\maketitle

\section{Introduction}
\label{sec:intro}

As suggested first by Jaffe
\cite{jaffe_multiquark_1977, *jaffe_multiquark_1977-1, *jaffe_color_2000,
*jaffe_diquarks_2003, *jaffe_exotica_2005},
it is most plausible that in QCD, the lightest scalar
mesons with $J^P = 0^+$ are composed not just of a quark and anti-quark,
but contain a significant admixture of tetraquark states,
with two quarks and two anti-quarks
\cite{black_mechanism_2000, fariborz_toy_2005, fariborz_two_2008, fariborz_note_2008, black_putative_1999, *fariborz_model_2007, *fariborz_global_2009, *fariborz_probing_2011, *fariborz_chiral_2011, *fariborz_chiral_2014, *fariborz_probing_2015, tornqvist_lightest_2002, *close_scalar_2002, *amsler_mesons_2004, * napsuciale_chiral_2004, *napsuciale_chiral_2004-3, *maiani_new_2004, *pelaez_light_2004, *pennington_can_2007, *t_hooft_theory_2008, *pelaez_chiral_2011,*heupel_tetraquark_2012, *mukherjee_low-lying_2012, *chen_1_2015, *eichmann_light_2016, giacosa_mixing_2007, parganlija_vacuum_2010, *giacosa_spontaneous_2010, *janowski_glueball_2011, *parganlija_meson_2013, *wolkanowski_scalar-isovector_2014, *janowski_is_2014, *ghalenovi_masses_2015, *giacosa_mesons_2016, olive_review_2014-1, *pelaez_controversy_2015}.
Recently, there is increasing experimental evidence 
for tetraquark and even pentaquark states of heavy quarks
\cite{karliner_doubly_2014, *stone_pentaquarks_2015}.

In this paper we concentrate on light quarks,
and generalize the standard analysis of the chiral phase
transition at nonzero temperature
\cite{goldberg_chiral_1983, pisarski_remarks_1984, pelissetto_relevance_2013, *nakayama_approaching_2014, *nakayama_bootstrapping_2015, meggiolaro_remarks_2013, *fejos_fluctuation_2014, *fejos_functional_2015, *sato_linking_2015, *eser_functional_2015}
to consider how tetraquarks can affect the chiral transition
\cite{heinz_role_2009, *gallas_nuclear_2011, *heinz_chiral_2015, mukherjee_chiral_2014}.
We limit ourselves to three colors, and
start with the case of two flavors, showing that tetraquarks probably
have a small effect on the chiral phase transition.  For three flavors,
though, if the quarks are sufficiently light then it is
possible --- although not guaranteed --- that the tetraquark field
generates a {\it second} chiral phase transition.
In the chiral limit both chiral phase transitions are of first order.
We discuss implications for the phase diagram of QCD at
nonzero temperature and chemical potential, and
conclude with some speculations about four flavors. 

A detailed comparison of models with tetraquarks 
to the hadronic spectrum is necessarily complicated, and 
involves not just the masses of hadronic states, but their decays
\cite{jaffe_multiquark_1977, *jaffe_multiquark_1977-1, *jaffe_color_2000,
*jaffe_diquarks_2003, *jaffe_exotica_2005, black_mechanism_2000, fariborz_toy_2005, fariborz_two_2008, fariborz_note_2008, black_putative_1999, *fariborz_model_2007, *fariborz_global_2009, *fariborz_probing_2011, *fariborz_chiral_2011, *fariborz_chiral_2014, *fariborz_probing_2015, tornqvist_lightest_2002, *close_scalar_2002, *amsler_mesons_2004, * napsuciale_chiral_2004, *napsuciale_chiral_2004-3, *maiani_new_2004, *pelaez_light_2004, *pennington_can_2007, *t_hooft_theory_2008, *pelaez_chiral_2011,*heupel_tetraquark_2012, *mukherjee_low-lying_2012, *chen_1_2015, *eichmann_light_2016, giacosa_mixing_2007, parganlija_vacuum_2010, *giacosa_spontaneous_2010, *janowski_glueball_2011, *parganlija_meson_2013, *wolkanowski_scalar-isovector_2014, *janowski_is_2014, *ghalenovi_masses_2015, *giacosa_mesons_2016, olive_review_2014-1, *pelaez_controversy_2015}.
Thus our discussion is largely qualitative, to emphasize what we
find is an unexpected relation between hadronic phenomenology at
zero temperature and the phase transitions of QCD.

\section{Notation}
\label{sec:standard}

Left and right handed quarks and anti-quarks are defined as
\begin{equation}
q_{L,R} = {\cal P}_{L,R} \; q \;\; ; \;\;
\overline{q}_{L,R} = \overline{q} \; {\cal P}_{R,L} 
\;\; ; \;\; 
{\cal P}_{L,R}
= \frac{1 \pm \gamma_5}{2} \; ,
\end{equation}
with $\gamma_5^2 = 1$.  

We assume there are $N_f$ flavors of massless quarks, which transform
under the chiral symmetry group of 
$SU(N_f)_L \times SU(N_f)_R \times U(1)_A$ as
\begin{equation}
q_L \rightarrow {\rm e}^{- i \alpha/2} \; U_L \, q_L \;\;\; ; \;\;\;
\overline{q}_L \rightarrow {\rm e}^{+ i \alpha/2} \; \overline{q}_L \, 
U_L^\dagger \;\;\; ; \;\;\;
q_R \rightarrow {\rm e}^{+ i \alpha/2} \; U_R \, q_R \;\;\; ; \;\;\;
\overline{q}_R \rightarrow {\rm e}^{- i \alpha/2} \; 
\overline{q}_R \, U_R^\dagger \; ;
\end{equation}
$U_{L,R}$ are elements of $SU(N_f)_{L,R}$ and 
$\alpha$ is an axial rotation in $U(1)_A$.  

For most of our discussion we implicitly limit ourselves to the case
of nonzero temperature and zero quark chemical potential.  This
allows us to assume that the $U(1)$ symmetry for quark number
remains unbroken.
At nonzero chemical potential color superconductivity
can occur, which spontaneously breaks this $U(1)$ symmetry
\cite{alford_color-flavor_1999, *schaefer_continuity_1999, *pisarski_superfluidity_1999, *pisarski_why_1999, *pisarski_gaps_2000, rajagopal_condensed_2000, *alford_color_2008, pisarski_critical_2000}.
As discussed in Sec. (\ref{sec:phase_diagram_T_mu}), the generalization
to nonzero quark chemical potential requires a separate analysis.

To construct the effective fields it helps to explicitly denote
the flavor and color indices.  
The quark field $q^{aA}$, where $a= 1\ldots N_f$ 
is the flavor index for $N_f$ flavors, and $A=1\ldots N_c$ for $N_c$ colors.
The usual order parameter for chiral symmetry is given
by combining a left handed anti-quark and a right handed quark 
as a color singlet,
\begin{equation}
\Phi^{a b}
= \overline{q}^{\, bA}_L \; q^{a A}_R \; .
\label{eq:define_phi}
\end{equation}
This field transforms as $\overline{{\bold N}}_f \times {\bold N}_f$
under $SU(N_f)_{L}$ and $SU(N_f)_{R}$:
\begin{equation}
\Phi \rightarrow {\rm e}^{+ i \alpha} \; 
U_R \; \Phi \; U_L^\dagger
\; .
\label{eq:transform_Phi}
\end{equation}
Under the axial $U(1)_A$ symmetry we can choose the convention that $\Phi$
has charge $=+1$.

We note that the combination of anti-quark and a quark with the same
chirality automatically vanishes: {\it e.g.}, 
$\overline{q}_L q_L = \overline{q} \, {\cal P}_R {\cal P}_L \, q = 0$.
In contrast, for tetraquarks 
it is possible to pair two diquark fields of the same
chirality, Eq. (\ref{chiral_zeta_twofl}) and Sec. (\ref{sec:tetra_same})

The chirally invariant couplings of quarks to the gauge field $A_\mu$
and to the chiral field $\Phi$ are
\begin{equation}
{\cal L}^{qk}_{\Phi} = 
\overline{q}_L 
\!\not \!\! D \,
q_L
+ 
\overline{q}_R
\!\not \!\! D \,
q_R
+
y_\Phi
\left(
\overline{q}_R
\, \Phi\,
q_L
+
\overline{q}_L
\, \Phi^\dagger \,
q_R
\right)
\; ,
\label{eq:chiral_inv_lag}
\end{equation}
where $D_\mu = \partial_\mu - i g A_\mu$ is the covariant derivative.

The Yukawa term  $\sim y_\Phi$ which couples
quarks to the chiral field $\Phi$ is an effective coupling.
Including such a term is useful in constructing an
effective model for the chiral transition \cite{pisarski_chiral_2016}.
We only write this term in order to contrast the difference
between the possible effective
couplings between quarks and the tetraquark fields
in Eqs. (\ref{eq:quark_zeta_two}) and
(\ref{eq:quark_zeta_three}).

We note 
that it is possible for chiral symmetry to be broken not by a 
quark antiquark operator in the 
$\bold{\overline{3}} \times \bold{3}$ representation, 
but by a four quark operator in the 
$\bold{8} \times \bold{8}$ representation
\cite{stern_two_1998, kogan_chiral_1998}.   These four quark operators differ
from the tetraquark operators which we consider.  In QCD,
though, there are general arguments against this possibility
\cite{kogan_chiral_1998}, and certainly no indication
from numerical simulations on the lattice that this occurs
\cite{cheng_transition_2006, *bazavov_equation_2009, *cheng_qcd_2008, *fodor_phase_2009, *aoki_qcd_2009, *borsanyi_qcd_2010, *borsanyi_is_2010, *cheng_equation_2010, *bazavov_chiral_2012, *bhattacharya_qcd_2014, *bazavov_equation_2014, *borsanyi_full_2014, buchoff_qcd_2014, ding_thermodynamics_2015, *ratti_lattice_2016}.

\section{Two Flavors}
\label{sec:tetra_two}

The most attractive channel for the scattering of two quarks is antisymmetric
in both flavor and color
\cite{jaffe_multiquark_1977, *jaffe_multiquark_1977-1, *jaffe_color_2000,
*jaffe_diquarks_2003, *jaffe_exotica_2005}.  
For two flavors, a diquark in this channel is then
an anti-triplet in color and an isosinglet in flavor,
\begin{equation}
\chi_L^A =
\epsilon^{A B C} \;\epsilon^{a b} \; 
(q_L^{a B})^T \; {\cal C}^{-1} \; q_L^{b \, C} \;  ,
\label{eq:diquark_twoflavor}
\end{equation}
where ${\cal C}$ is the charge conjugation matrix \cite{fariborz_note_2008}.  
In a basis
where $\gamma_5 = ({\bf 1}_2,-{\bf 1}_2)$ is diagonal,
${\cal C} = {\rm diag}(- \sigma_2,\sigma_2)$.  The transpose
of the quark field and the charge
conjugation matrix ${\cal C}$ are necessary to form a Lorentz
scalar.  This combination is
naturally related to the diquark condensates for color
superconductivity
\cite{alford_color-flavor_1999, *schaefer_continuity_1999, *pisarski_superfluidity_1999, *pisarski_why_1999, *pisarski_gaps_2000, rajagopal_condensed_2000, *alford_color_2008, pisarski_critical_2000}.

To obtain a spin zero field we combine
left handed diquark with a right handed diquark to form
\begin{equation}
\zeta =
(\chi_R^A)^* \; \chi_L^A \; .
\label{eq:define_zeta_two}
\end{equation}
The tetraquark field $\zeta$ is a color singlet and complex valued.
It is invariant under $SU(2)_L \times SU(2)_R$, but
transforms under axial $U(1)_A$ as
\begin{equation}
\zeta \rightarrow {\rm e}^{- 2 i \alpha} \; \zeta \; ,
\label{eq:charge_zeta_twofl}
\end{equation}
so that $\zeta$ has axial $U(1)_A$ charge $= - 2$.

Unlike for $\Phi$, we can also form tetraquark fields from diquarks of
the {\it same} chirality: 
\begin{equation}
\zeta_L = (\chi_L^A)^* \; \chi_L^A \;\;\; , \;\;\;
\zeta_R = (\chi_R^A)^* \; \chi_R^A
\; .
\label{chiral_zeta_twofl}
\end{equation}
Both $\zeta_L$ and $\zeta_R$
are real valued and singlets under all flavor transformations.
Thus while they can be constructed, there is no reason to expect that
they should significantly affect the dynamics in any interesting way.
In particular, they appear through terms which are linear in
themselves, and have an expectation value at any temperature.  

We thus turn to constructing an effective Lagrangian which couples
the usual chiral field $\Phi$ and the tetraquark field $\zeta$
under an exact chiral symmetry of $SU(2)_L \times SU(2)_R$.

We assume that in counting mass dimensions, all scalar
fields have mass dimension one, as holds
for a fundamental scalar in four spacetime dimensions.  
Since the quarks have mass dimension $3/2$, this is different
from their nominal mass dimension, which is three for $\Phi$,
and six for $\zeta$.  This is, however, a standard
assumption in constructing effective models, and is certainly justified 
by the renormalization group near
a transition of second order.  We then catagorize all terms up to
quartic order in $\Phi$ and $\zeta$.

While in the chiral limit the $SU(2)_L \times SU(2)_R$ symmetry is exact, 
the axial $U(1)_A$ symmetry is only valid classically, and is spontaneously
broken quantum-mechanically by 
topologically nontrivial configurations such as instantons
\cite{gross_qcd_1981, t_hooft_how_1986}.
There still persists a discrete axial symmetry of $Z(2)_A$.
The simplest operator which is invariant under $SU(2)_L \times SU(2)_R$,
but not $U(1)_A$, is the determinant of $\Phi$.
For two flavors, under axial $U(1)_A$ this operator has
axial charge $= +2$,
\begin{equation}
{\rm det} \Phi \rightarrow {\rm e}^{2 i \alpha} \; {\rm det} \Phi
\; .
\end{equation}
This is invariant if $\alpha = 0$ or $\pi$, which is the 
residual symmetry of axial $Z(2)_A$.  

Consequently, any couplings which invariant under $Z(2)_A$ but not
$U(1)_A$ are 
nonzero in vacuum and for a range of temperature.  Eventually,
at high temperature the breaking of axial $U(1)_A$ is
only due to instantons.  This is suppressed by a high power of
temperature \cite{gross_qcd_1981}, 
so that axial $U(1)_A$ is effectively restored
as the temperature $T \rightarrow \infty$.
This is supported by numerical simulations on the
lattice \cite{borsanyi_axion_2016}.

To help catagorize the possible terms in effective potentials
it helps to start with
those which persist at high temperature, where axial $U(1)_A$
is an approximate symmetry.  The $U(1)_A$ invariant terms that
only involve $\Phi$ are
\begin{equation}
{\cal V}_{\Phi}^\infty
= m_\Phi^2 \; {\rm tr} \left( \Phi^\dagger \Phi \right)
+ \lambda_{\Phi 1} \, {\rm tr} \left(\Phi^\dagger \Phi\right)^2
+ \lambda_{\Phi 2} \, \left({\rm tr} \Phi^\dagger \Phi\right)^2
\; .
\label{eq:phi_two_inf}
\end{equation}
These terms are standard in linear sigma models.
For two flavors, $|{\rm det}\Phi|^2$ is also a quartic term,
but because $\Phi^\dagger \Phi$ is a Hermitian matrix,
this can be expressed as a sum of the two terms above,
$|{\rm det}\Phi|^2 = 
{\rm det}\Phi^\dagger \Phi = (({\rm tr}\Phi^\dagger \Phi)^2
- {\rm tr}(\Phi^\dagger \Phi)^2)/2$.

There are two $U(1)_A$ invariant terms which only involve $\zeta$,
\begin{equation}
{\cal V}_{\zeta}^{\infty} = 
m_\zeta^2 \; |\zeta|^2
+ \lambda_\zeta \; (|\zeta|^2)^2
\; .
\label{eq:zeta_twofl_inf}
\end{equation}

Lastly, there are two $U(1)_A$ invariant terms coupling
$\Phi$ and $\zeta$,
\begin{equation}
{\cal V}_{\zeta \Phi}^\infty = 
+ \kappa_\infty \, \left( \zeta \; {\rm det} \Phi + {\rm c.c.} \right)
+ \lambda_{\zeta \phi 1} \, |\zeta|^2 
\; {\rm tr} \left( \Phi^\dagger \Phi \right)
\; .
\label{eq:phi_zeta_twofl_u1}
\end{equation}
The last term is unremarkable, as both $|\zeta|^2$ and 
${\rm tr} ( \Phi^\dagger \Phi)$ are each separately
invariant under $U(1)_A$.
The first term, however, is novel: it is
a trilinear coupling between one $\zeta$ field, with axial
charge $= -2$, and two $\Phi$'s, with charge $= +1$.  
Adding the complex conjugate (c.c.) assures the total term is real.
There is an analogous term for three flavors, 
Eq. (\ref{trilinear_u1_sym_three}).  

We then move on to catagorize the full set of terms which contribute
at finite temperature, where the $U(1)_A$ symmetry is reduced to
$Z(2)_A$.  
There are three terms involving only $\Phi$:
\begin{equation}
{\cal V}_{\Phi}^{A}
=
\kappa_{\Phi} \, \left( {\rm det} \Phi + {\rm c.c.} \right)
+ \lambda_{\Phi 3} \, \left( {\rm det} \Phi + {\rm c.c.} \right) 
{\rm tr} \left( \Phi^\dagger \Phi \right)
+ \lambda_{\Phi 4} \, \left( {\rm det} \Phi  + {\rm c.c.} \right)^2
\; .
\label{eq:phi_zeta_twofl_z2}
\end{equation}
The first is a mass term makes the $\eta$ meson heavy,
and so splits the $U(1)_A$ symmetry in the spectrum
\cite{t_hooft_how_1986}.  The other two are couplings of quartic order.
Since $\Phi$ itself is not a Hermitian matrix, 
${\rm det}\Phi$ does not reduce to traces of $\Phi$,
and these are new, independent couplings.  (This is can be
checked by taking the elements of $\Phi$ to be only off-diagonal
and complex.)

At zero temperature, the tetraquark field $\zeta$ is a $Z(2)_A$ singlet,
and so there is no symmetry relating the
real and imaginary parts of $\zeta$.  The real part of $\zeta$,
$\zeta_r$, is even under parity, while the imaginary part, $\zeta_i$,
is odd.  We start with the terms for $\zeta_r$.   As it
is a $Z(2)_A$ singlet and parity even, the couplings of $\zeta_r$ with itself
involves {\it arbitrary} powers:
\begin{equation}
{\cal V}_{\zeta_r}^A = 
h_r \, \zeta_r
+ m_r^2 \, \zeta_r^2
+ \kappa_r \, \zeta_r^3
+ \lambda_r \, \zeta_r^4 
\; .
\label{eq:zeta_twofl}
\end{equation}

Assuming that the underlying theory, such as QCD, does not spontaneously
break parity, then only even powers of
the imaginary part $\zeta_i$ can appear.  Otherwise,
arbitrary combinations of $\zeta_r$ and $\zeta_i^2$ enter:
\begin{equation}
{\cal V}_{\zeta_i}^A = 
+ m_i^2 \, \zeta_i^2
+ \kappa_{i} \, \zeta_r \, \zeta_i^2
+ \lambda_{i1} \, \zeta_i^4 
+ \lambda_{i2} \, \zeta_r^2 \, \zeta_i^2 
\; .
\label{eq:zeta_twofl_im}
\end{equation}
However, $\zeta_i$ does not play a significant role in the chiral
phase transition, and so we neglect it henceforth.

That leaves couplings between $\zeta_r$ and $\Phi$,
\begin{equation}
{\cal V}_{\zeta \Phi}^A
= \kappa_{\zeta \Phi } 
\; \zeta_r \; {\rm tr} \left( \Phi^\dagger \Phi \right)
+ \lambda_{\zeta \Phi 2} 
\, \zeta_r^2 \, \left( {\rm det}\, \Phi + {\rm c.c.} \right)
\; .
\label{eq:zetaPhi_two_A}
\end{equation}
The trilinear coupling between $\zeta_r$ and ${\rm tr} (\Phi^\dagger \Phi)$
was noticed first by Giacosa \cite{giacosa_mixing_2007}.

These effective Lagrangians can be used to analyze the effect
of the tetraquark field $\zeta$ on the chiral phase transition.
At zero temperature we assume that the chiral symmetry is 
spontaneously broken, with $\langle \Phi \rangle \neq 0$.  
Further, since there is no reason why $h_r$ in Eq. (\ref{eq:zeta_twofl})
should vanish, we also assume that 
$\langle \zeta_r \rangle \neq 0$ at $T=0$.

As the temperature changes
all of the $U(1)_A$ invariant couplings in 
Eqs. (\ref{eq:phi_two_inf}), (\ref{eq:zeta_twofl_inf}), and
(\ref{eq:phi_zeta_twofl_u1}) are nonzero at any $T$.
In contrast, the $Z(2)_A$ invariant couplings in
Eqs. (\ref{eq:phi_zeta_twofl_z2}), (\ref{eq:zeta_twofl}),
(\ref{eq:zeta_twofl_im}), and 
(\ref{eq:zetaPhi_two_A}) vanish as $T \rightarrow \infty$.

In particular, while $\zeta_r$ and $\zeta_i$ are not related at
zero temperature, as $T\rightarrow \infty$, we should have the
(approximate) restoration of axial $U(1)_A$ symmetry.  This implies
that $\zeta_r$ and $\zeta_i$ are degenerate, with
$\langle \zeta_r \rangle \rightarrow 0$ as $T \rightarrow \infty$.

A chiral phase transition occurs when the expectation value
of $\Phi$ vanishes.  
We first review the standard picture in the absence of 
the tetraquark field $\zeta$.
If axial $U(1)_A$ is badly broken at $T_\chi$,
then the chiral symmetry is $SU(2)_L \times SU(2)_R$.
Assuming that quartic couplings are positive at $T_\chi$,
when $m_\Phi^2$ vanishes there is a second order phase transition in the
universality class of $O(4)$ symmetry.  If $Z(2)_A$ is approximately
$U(1)_A$ by $T_\chi$
then the chiral transition could be induced to be first order through
fluctuations \cite{pisarski_remarks_1984}.  Another possibility is that 
$SU(2)_L \times SU(2)_R \times U(1)_A = O(4) \times O(2)$
has an infrared stable fixed point in a new universality class
\cite{pelissetto_relevance_2013, *nakayama_approaching_2014, *nakayama_bootstrapping_2015}.  For analyses in effective models, see Ref.
\cite{meggiolaro_remarks_2013, *fejos_fluctuation_2014, *fejos_functional_2015, *sato_linking_2015, *eser_functional_2015}.

Including the tetraquark field $\zeta$ does not appear to
significantly affect the chiral phase transition.  
For two flavors, {\it all} of the mixing terms between
$\zeta$ and $\Phi$,
Eqs. (\ref{eq:phi_zeta_twofl_u1}) and (\ref{eq:zetaPhi_two_A}),
are quadratic in $\Phi$.  Consequently, 
the mixing between $\Phi$ and $\zeta$ is $\sim \langle \Phi \rangle$.
If the chiral transition is of second order, at $T_\chi$ this
mixing vanishes, and only $\Phi$ is a critical field.  Both
$\zeta_r$ and $\zeta_i$ are massive fields which mix with $\Phi$
due to cubic terms.  

If the chiral transition for two flavors
is of first order, then of course both $\langle \Phi \rangle$ and
$\langle \zeta_r \rangle$ have a discontinuity at $T_\chi$.

Given the generality of the potentials, it is possible
that there is a phase transition associated with $\zeta_r$,
independent of that for $\Phi$.
Even if $h_r$ vanishes at one given temperature, due to the cubic
terms in $\zeta_r$, $\langle \zeta_r \rangle$ should still be nonzero.
As noted by 
Mukherjee and Huang \cite{mukherjee_chiral_2014},
this does not exclude the possibility of a first order transition
at which $\langle \zeta_r \rangle$ jumps discontinuously from one
value to another.  While possible, however, there is 
compelling reason why such a first order transition in $\zeta_r$ 
should occur.  

We briefly discuss the mass spectrum of the model.
As a complex valued field, $\Phi$ has components with $J^P = 0^+$ and
$0^-$.  The $0^+$ is composed of an isosinglet, 
the $\sigma$, and and isotriplet, analogous to the $\vec{a}_0$.  For the
$0^-$ part we have an isosinglet $\eta$ and 
an isotriplet of pions, $\vec{\pi}$.  In addition,
$\zeta$ contains two isosinglet fields, $\zeta_{r,i}$ with
$J^P = 0^{\pm}$.  

At zero temperature, $\langle \Phi \rangle \neq 0$ generates a massless
pion and $\eta$, and a massive $\sigma$ and $\vec{a}_0$.  Terms
which are only invariant under $Z(2)_A$ make the $\eta$ massive,
pushing the mass of the $\sigma$ down \cite{t_hooft_how_1986}.
With the tetraquark field, all that happens is that
the $\zeta_r$ field mixes with the $\sigma$, as does
the $\zeta_i$ field with the $\eta$; the isotriplet states are unaffected.

As noted above, the mixing between $\Phi$ and $\zeta$ is
$\sim \langle \Phi \rangle$, and so vanishes in
the chirally symmetric phase.  At very high temperatures where
$U(1)_A$ symmetry is approximately valid,
the $\Phi$ multiplet is (nearly) degenerate, as are $\zeta_r$ and $\zeta_i$.
There is no reason why the masses of $\Phi$ and $\zeta$ should be
related to one another, although the two fields couple 
through Eq. (\ref{eq:phi_zeta_twofl_u1}).

We conclude by noting that because the tetraquark field is a singlet
under $SU(2)_L \times SU(2)_R$, there is no Yukawa coupling analogous
that between $\Phi$ and the quark fields, Eq. (\ref{eq:chiral_inv_lag}).
There is, however, a $U(1)_A$ invariant coupling,
\begin{equation}
y_{\zeta 2}
\left( \; (\chi_L^A)^* \; \zeta \; \chi_R^A 
+ \; \; (\chi_R^A)^* \; \zeta^* \; \chi_L^A \;
\right)
\; ,
\label{eq:quark_zeta_two}
\end{equation}
using $\chi_{L,R}^A$ from Eq. (\ref{eq:diquark_twoflavor}).
As each $\chi_{L,R}$ is a diquark operator, this is a coupling
between $\zeta$ and {\it four} quarks, so 
the coupling $y_{\zeta 2} \sim 1/{\rm mass}^3$.
A coupling with such a large, negative mass dimension is much
less important than those given above, which have either positive
or vanishing mass dimension.

\section{Three flavors}
\label{sec:tetra_three}

\subsection{Tetraquarks with opposite chirality}
\label{sec:tetra_three_general}

For three flavors the diquark field is 
\begin{equation}
\chi^{a A}_L =
\epsilon^{a b c} \; \epsilon^{A B C} \;
( q^{b B}_L )^T  \; {\cal C}^{-1} \; q^{c C}_L \;  .
\label{eq:diquark_threeflavor}
\end{equation}
Because of the anti-symmetric tensor, $\chi_L$ transforms
as an anti-triplet, $\overline{{\bold 3}}$, in both color and flavor.  
The diquark fields $\chi_L$ and $\chi_R$ can be combined into a color singlet,
tetraquark field,
\begin{equation}
\zeta^{a b} =
(\chi_R^{a A})^* \; \chi_L^{b A} \; .
\label{eq:define_zeta_three8}
\end{equation}
Unlike two flavors, $\zeta$ transforms nontrivially
under the $SU(3)_L \times SU(3)_R$ chiral symmetry:
\begin{equation}
\zeta \rightarrow
{\rm e}^{- 2 i \alpha} \; U_R \, \zeta \, U_L^\dagger \; .
\label{eq:define_zeta_three_transf}
\end{equation}
Note that while we define 
$\Phi \sim \overline{q}_L q_R$, as a left-right field
Eq. (\ref{eq:define_phi}), we choose to define 
$\zeta \sim \chi_R^\dagger \chi_L \sim \overline{q}_R \overline{q}_R
q_L q_L$ as right-left.  We do this so that both $\zeta$ and $\Phi$
in the same way under $SU(3)_L \times SU(3)_R$, 
as $\bold{\overline{3}} \times \bold{3}$.
Because of this difference, they have opposite
signs under the axial $U(1)_A$ symmetry:
$\Phi^{a b}$ has axial charge $+1$, while $\zeta^{a b}$ has charge $-2$.

As for two flavors, we first catagorize the interactions which
are $U(1)_A$ invariant.  Those involving just $\Phi$ are
\begin{equation}
{\cal V}_{\Phi}^\infty
= m_\Phi^2 \; {\rm tr} \left( \Phi^\dagger \Phi \right)
+ \lambda_{\Phi 1} \, {\rm tr} \left(\Phi^\dagger \Phi\right)^2
+ \lambda_{\Phi 2} \, \left( {\rm tr} \left(\Phi^\dagger \Phi\right)
\right)^2
\; ,
\label{eq:phi_pure_three}
\end{equation}
and similarly for $\zeta$, 
\begin{equation}
{\cal V}_{\zeta}^\infty
= m_\zeta^2 \; {\rm tr} \left( \zeta^\dagger \zeta \right)
+ \lambda_{\zeta 1} \, {\rm tr} \left(\zeta^\dagger \zeta\right)^2
+ \lambda_{\zeta 2} \, 
\left( 
{\rm tr} \left(\zeta^\dagger \zeta\right)
\right)^2
\; .
\label{eq:zeta_pure_three}
\end{equation}

Even under the assumption of $U(1)_A$ symmetry, there are numerous
couplings between $\zeta$ and $\Phi$.  The most interesting is
a trilinear coupling between $\zeta$ and $\Phi$,
\begin{equation}
{\cal V}_{\zeta \Phi,3}^\infty
= \kappa_\infty \; \epsilon^{a b c} \; \epsilon^{a' b' c'}\;
\left(
\zeta^{a a'} \, \Phi^{b b'} \, \Phi^{c c'}
+ {\rm c.c.}
\right)
\; .
\label{trilinear_u1_sym_three}
\end{equation}
This term ties left handed indices with left handed, and right with right,
and so is
invariant under $SU(3)_L \times SU(3)_R$.  
(Note that both $\Phi^{a b}$ and $\zeta^{a b}$ are defined so
that the first
index is for $SU(3)_R$, and the second for $SU(3)_L$.)
This is is invariant 
under the axial $U(1)_A$ symmetry because there is one
$\zeta$ with charge $-2$ and two $\Phi$'s with charge $+1$.  
This coupling is analogous
to that for two flavors, $\sim \kappa_\infty \, \zeta \, {\rm det} \Phi$ in 
Eq. (\ref{eq:phi_zeta_twofl_u1}).  

There are four quartic couplings which are invariant under $U(1)_A$
and mix $\zeta$ and $\Phi$:
\begin{eqnarray}
{\cal V}_{\zeta \Phi,4}^\infty
&=& 
 \lambda_{\zeta \Phi 1} \, {\rm tr} 
\left(\zeta^\dagger \, \zeta \,
\Phi^\dagger \, \Phi 
\right)
+ \lambda_{\zeta \Phi 2} \,
{\rm tr} 
\left(\zeta^\dagger \, \Phi \, \Phi^\dagger \, \zeta  \right)
\nonumber \\
\;\; &+& \lambda_{\zeta \Phi 3} \, 
{\rm tr} 
\left(\zeta^\dagger \, \zeta \right)
{\rm tr}
\left( \Phi^\dagger \, \Phi \right)
+ \lambda_{\zeta \Phi 4} \,
{\rm tr} 
\left(\zeta^\dagger \, \Phi \right)
{\rm tr} 
\left(\Phi^\dagger \, \zeta \right)
\; .
\label{quartic_u1_sym_threefl}
\end{eqnarray}

We next turn to terms which are invariant only under $Z(3)_A$
and not $U(1)_A$.  The most
important was noted first by Black, Fariborz, and Schechter 
\cite{black_mechanism_2000}.  
This is a quadratic term, which {\it directly} mixes $\zeta$ and $\Phi$,
\begin{equation}
{\cal V}_{\zeta\Phi,2}^A
= \; m_{\zeta \Phi}^2 \; {\rm tr} 
\left(
\zeta^\dagger  \Phi 
+ \Phi^\dagger \zeta
\right)
\; .
\label{eq:mix_zeta_phi_three}
\end{equation}
This has axial charge $\pm 3$ and so is $Z(3)_A$ invariant.
The existence of
this mixing term is 
an immediate consequence of the fact that $\zeta$ and $\Phi$
transform in the same representation of the chiral symmetry group.

There are three cubic terms which are $Z(3)_A$ invariant,
\begin{equation}
{\cal V}_{\zeta\Phi,3}^A
=
\kappa_\Phi \, \left(
{\rm det}\Phi
+ {\rm c.c.}
\right)
+ \kappa_\zeta \,
\left( 
{\rm det}\zeta
+ {\rm c.c.}
\right)
+ \, \kappa_{\zeta \Phi} \; \epsilon^{a b c} \, \epsilon^{a' b' c'}
\left(
\zeta^{a a'} \, \zeta^{b b'} \, \Phi^{c c'}
+ {\rm c.c.}
\right)
\; .
\label{trilinear_z3}
\end{equation}
The last term
is clearly similar to that in 
Eq. (\ref{trilinear_u1_sym_three}), except that it involves
two $\zeta$'s and one $\Phi$, with axial charge $\mp 3$.  

There are six quartic terms which are $Z(3)_A$ invariant,
\begin{eqnarray}
{\cal V}_{\zeta \Phi,4}^A &=&
\lambda_{\zeta \Phi 5} \,
\left(
{\rm tr} 
\left(
\zeta^\dagger \zeta \zeta^\dagger \Phi
\right)
+ {\rm c.c.}
\right)
+ \lambda_{\zeta \Phi 6} \,
\left(
{\rm tr} 
\left(\zeta^\dagger \Phi \right)^2
+ {\rm c.c.}
\right)
+ \lambda_{\zeta \Phi 7} \,
\left(
{\rm tr} 
\left(
\zeta^\dagger \, \Phi \, \Phi^\dagger \, \Phi
\right)
+ {\rm c.c.}
\right)
\nonumber \\
&+& 
\lambda_{\zeta \Phi 8} \,
\left(
{\rm tr} 
\left(\zeta^\dagger \, \zeta \right)
{\rm tr} 
\left(\zeta^\dagger \, \Phi \right)
+ {\rm c.c.}
\right)
+ \lambda_{\zeta \Phi 9} \,
\left(
\left(
{\rm tr} 
\left(\zeta^\dagger \, \Phi \right)
\right)^2 
+ {\rm c.c.}
\right)
\nonumber \\
&+& 
\lambda_{\zeta \Phi 10} \,
\left(
{\rm tr} 
\left(\zeta^\dagger \, \Phi \right)
{\rm tr} 
\left(\Phi^\dagger \, \Phi \right)
+ {\rm c.c.}
\right)
\; .
\label{eq:quartic_zeta_phi_z3}
\end{eqnarray}
These terms agree with 
Fariborz, Jora, and Schechter, Appendix A in
Refs. \cite{fariborz_toy_2005} and \cite{fariborz_two_2008}.

As discussed before for two flavors, we assume that all couplings
which are invariant under $Z(3)_A$ but not $U(1)_A$ are large
at zero temperature, but negligible at high temperature.  We
do not assume that they are small at the chiral phase transition.

At zero temperature we expect that the chiral symmetry is broken
by a nonzero expectation value for $\Phi$, $\langle \Phi \rangle \neq 0$.
Since we take
the chiral symmetry to be exact, $\langle \Phi \rangle$ is proportional
to the unit matrix.  Because of the mixing term in Eq.
(\ref{eq:mix_zeta_phi_three}), an expectation value for
$\Phi$ automatically induces one 
for $\zeta$, with $\langle \zeta \rangle \neq 0$.

At high temperature we expect the chiral symmetry is restored, so
$\langle \Phi \rangle = \langle \zeta \rangle = 0$.  
Further,
because the direct mixing between $\Phi$ and $\zeta$,
Eq. (\ref{eq:mix_zeta_phi_three}), which is only invariant 
under $Z(3)_A$, at high temperature
the masses of $\zeta$ and $\Phi$ do not mix.  The fields do interact
through $U(1)_A$ invariant couplings such as 
Eqs. (\ref{trilinear_u1_sym_three}) and (\ref{quartic_u1_sym_threefl}).

The interesting question is how chiral symmetry is restored.
This depends upon the details of the effective Lagrangian.  For example,
assume that $m_\zeta^2$ is {\it very} large and positive at zero temperature.
Then an expectation value of $\zeta$ is induced only by its mixing
with $\Phi$: 
the phase transition is driven by the interactions
of $\Phi$ with itself, and $\zeta$ plays a tangential role.

Since both $\Phi$ and $\zeta$ lie in the same representation
of $SU(3)_L \times SU(3)_R$, the converse is also possible: if
$m_\Phi^2$ is large and positive at zero temperature, then chiral
symmetry breaking and restoration is driven by the tetraquark field,
$\zeta$.  

We suggest that it is possible that
both $\Phi$ and $\zeta$ play important
roles in the breaking of chiral symmetry at zero temperature,
and its restoration at $T_\chi$.

If so, then it is very possible that there are {\it two} chiral
phase transitions, at temperatures
$T_{\widetilde{\chi}}$ and $T_\chi$, where $T_{\widetilde{\chi}} < T_\chi$.
Because of the cubic terms in $\zeta$ and $\Phi$, 
Eqs. (\ref{trilinear_u1_sym_three}) and (\ref{trilinear_z3}),
both transitions are presumably of first order.
As the temperature increases from zero, there is first 
a phase transition at $T_{\widetilde{\chi}}$, where both $\langle \Phi \rangle$ and
$\langle \zeta \rangle$ jump discontinuously.  Because
of their mixing, both condensates remain nonzero above 
but close to $T_{\widetilde{\chi}}$.
As the temperature continues to increase, they 
jump again at $T_\chi$, and vanish for $T> T_\chi$.
Thus $T_\chi$ is properly termed the temperature for the restoration
of chiral symmetry.  Nevertheless, the transition at $T_{\widetilde{\chi}}$ is
also a chiral phase transition, since both expectation values jump
there.  It is simply not a transition above which the chiral symmetry
is restored.  

In terms of the effective Lagrangian,
there is a wide range of parameters in which there are two chiral
phase transitions.  The most obvious is if the mass squared of
both $\Phi$ and $\zeta$ are negative at zero temperature.  Then
given the bounty of cubic terms, it is extremely {\it un}natural 
for there to be only one phase transition.  

What we are suggesting is actually rather elementary.  If both
the usual chiral field $\Phi$ and the
tetraquark field $\zeta$ matter at zero
temperature, as suggested by hadronic phenomenology, then because
they lie in the {\it same} representation of $SU(3)_L \times SU(3)_R$,
it is very plausible that each chiral field drives a 
phase transition.

As shown by our discussion of two flavors, our conclusion is
special to three flavors.  As we discuss in
Sec. (\ref{sec:four_flavors}), even for four flavors the relevant
fields may differ, and be hexaquarks instead of tetraquarks.  

The importance of tetraquarks for the chiral phase transition is
also special to being close to the chiral limit.  For physical
values of the quark masses, numerical simulations on the lattice
find only one chiral phase transition
\cite{cheng_transition_2006, *bazavov_equation_2009, *cheng_qcd_2008, *fodor_phase_2009, *aoki_qcd_2009, *borsanyi_qcd_2010, *borsanyi_is_2010, *cheng_equation_2010, *bazavov_chiral_2012, *bhattacharya_qcd_2014, *bazavov_equation_2014, *borsanyi_full_2014, buchoff_qcd_2014};
for recent reviews, see 
\cite{ding_thermodynamics_2015, *ratti_lattice_2016}.  
As we argue in the next
section, the tetraquark field becomes more important as the quarks
become lighter.

We conclude this section by noting that unlike the case of two flavors,
that the tetraquark field can couple directly to quarks through a Yukawa
interaction similar to that for $\Phi$ in Eq. (\ref{eq:chiral_inv_lag}),
\begin{equation}
y_{\zeta 3}
\left(
\; \overline{q}_R
\, \zeta\,
q_L
+
\; \overline{q}_L
\, \zeta^\dagger \,
q_R
\;
\right)
\; .
\label{eq:quark_zeta_three}
\end{equation}
However, this coupling has axial charge $\mp 3$, and so
is invariant under $Z(3)_A$, but not $U(1)_A$.  Thus $y_{\zeta 3}$ 
vanishes as $T\rightarrow \infty$.

\subsection{Tetraquarks with the same chirality}
\label{sec:tetra_same}

Analogous to the case of two flavors, 
Eq. (\ref{chiral_zeta_twofl}), 
it is also possible to combine two diquark fields
with the same chirality:
\begin{equation}
\zeta_{L}^{a b} = \chi_{L}^{a A} \, (\chi_{L}^{b A})^* 
\;\;\; , \;\;\;
\zeta_{R}^{a b} = \chi_{R}^{a A} \, (\chi_{R}^{b A})^* 
\; .
\end{equation}
By their definition these fields are Hermitian, 
$\zeta_{L}^\dagger = \zeta_{L}$ and
$\zeta_{R}^\dagger = \zeta_{R}$.  
They transform as an adjoint field under the associated
flavor group, with axial charge zero:
\begin{equation}
\zeta_{L} \rightarrow U_{L}\, \zeta_{L} \, U_{L}^\dagger
\;\;\; , \;\;\;
\zeta_{R} \rightarrow U_{R} \, \zeta_{R} \, U_{R}^\dagger
\; .
\end{equation}
For the left handed fields,
their self interaction include
\begin{equation}
{\cal V}_{\zeta_L}
= h_{\zeta_L} \; {\rm tr} \left( \zeta_L \right)
+ m^2_{\zeta_l} {\rm tr} \left( \zeta_L^2 \right)
+ \kappa_{\zeta_L} \, {\rm tr} \left(\zeta_L^3 \right)
+ \lambda_{\zeta_L} \, {\rm tr} \left(\zeta_L^4 \right)
\; ;
\end{equation}
and similarly for $\zeta_R$.

The important point is that because they are Hermitian fields, and
carry zero charge under axial $U(1)_A$, then a term linear in 
either the trace of $\zeta_L$ or $\zeta_R$ is allowed at any temperature.
Thus we expect that each develops a nonzero expectation
value.  This is true at any temperature: 
even in the chirally
symmetric phase, if $\langle \zeta_{L} \rangle$ 
and $\langle \zeta_{R} \rangle$ are each proportional
to the unit matrix, then $SU(3)_L$ and $SU(3)_R$ symmetries remain
unbroken by these expectation values.  We assume this remains
valid at any temperature.

Couplings of the left and right handed tetraquark fields with $\Phi$ include
\begin{equation}
{\cal V}_{\zeta_L \Phi}
= 
\kappa_{\zeta_{L} \Phi}\; 
{\rm tr} \left( \zeta_L 
\, \Phi^\dagger \, \Phi \right)
+ 
\kappa_{\zeta_{R} \Phi}\; 
{\rm tr} \left( \zeta_R
\, \Phi \, \Phi^\dagger \right)
+ \lambda_{\zeta_{LR} \Phi}\; 
{\rm tr} \left( \zeta_L \, \Phi^\dagger \, \zeta_R \, \Phi \right) 
\; ;
\label{eq:zetaL_phi_three}
\end{equation}
plus other terms.   Invariance under parity requires
$\kappa_{\zeta_{L} \Phi} = \kappa_{\zeta_{R} \Phi}$.
There are, of course, also couplings with 
the left-right tetraquark field $\zeta$
as well as with $\Phi$.  If both $\zeta_L$ and $\zeta_R$ develop
expectation values which are proportional to the unit matrix, however, then
all of these terms reduce to couplings just between $\Phi$ and $\zeta$,
as written down previously.  For example, all of the terms
in Eq. (\ref{eq:zetaL_phi_three}) reduce just to 
${\rm tr} (\Phi^\dagger \Phi)$.
Consequently, we do not expect that whatever happens with $\zeta_L$
and $\zeta_R$ to materially affect the phase transitions in $\Phi$
and $\zeta$.  As for two flavors \cite{mukherjee_chiral_2014}, 
there can be first order transitions
associated with either field at any temperature, but there seems to be
no compelling dynamical reason for such transitions.

\section{Toy model}

In this section we discuss a simple model which illustrates
how two chiral phase transitions can arise for three massless flavors.

\subsection{Single chiral field}

We first review the chiral phase transition
for a single chiral field, $\Phi$.  Besides 
establishing notation, it helps to illustrate the range
of possible values.
We start with the Lagrangian of Eqs. (\ref{eq:phi_pure_three}) 
and (\ref{trilinear_z3}),
\begin{equation}
{\cal V}_{\Phi}(\Phi)
= m^2 \; {\rm tr} \left( \Phi^\dagger \Phi \right)
+ \kappa \, \left(
{\rm det} \Phi 
+ {\rm c.c.}
\right)
+ \lambda \, {\rm tr} \left(\Phi^\dagger \Phi\right)^2
\; .
\label{eq:toy_Phi}
\end{equation}
To avoid notational clutter, we drop the subscript $\Phi$, taking
$m^2_\Phi = m^2$, $\kappa_\Phi = \kappa$, and $\lambda_{\Phi 1} = \lambda$.
We also drop the coupling 
$\sim \lambda_{\Phi 2} ({\rm tr}( \Phi^\dagger \Phi))^2$.

In the chiral limit we take the expectation value of $\Phi$ to 
be diagonal,
\begin{equation}
\langle \Phi^{a b} \rangle = \phi \; \delta^{a b} 
\; .
\end{equation}
For this value,
\begin{equation}
{\cal V}_{\Phi}(\phi) = 
3 \, m^2 \, \phi^2 - \, 2\, \kappa \, \phi^3
+ 3 \, \lambda \, \phi^4 \; .
\end{equation}
The equation of motion for $\phi$ is
\begin{equation}
\frac{\partial {\cal V}_\Phi(\phi)}{\partial \phi}
= 6 \phi
\left(
m^2 - \, \kappa \, \phi
+ 2 \, \lambda \, \phi^2
\right)
 \; .
\label{eq:EOM_phi}
\end{equation}

In the chiral limit, there are only four distinct masses.
The fields with $J^P = 0^-$ are a degenerate octet,
composed of the pions, kaons, and the $\eta$, and a singlet $\eta'$.
Those with $J^P=0^+$ are a degenerate octet of the
$a_0$'s, $K^*$'s, and an $f_0$ meson, 
and a singlet $\sigma$ meson.

These four masses can be read off from Eqs. (68), (71), (77), and (81)
of Ref. \cite{pisarski_chiral_2016},
\begin{eqnarray}
m^2_{\pi} &=& m^2 - \, \kappa \, \phi +
2 \, \lambda \, \phi^2  \; ,
\nonumber \\
m^2_{\eta'} &=& m^2 + 2 \, \kappa \, \phi +
2 \, \lambda \, \phi^2  \; ,
\nonumber \\
m^2_{a_0} &=& m^2 + \kappa \, \phi +
6 \, \lambda \, \phi^2  \; ,
\nonumber \\
m^2_{\sigma} &=& m^2 - 2 \, \kappa \, \phi +
6 \, \lambda \, \phi^2  \; .
\label{eq:masses_sigma_toy}
\end{eqnarray}
The pion mass squared is proportional to the equation of motion,
Eq. (\ref{eq:EOM_phi}), and so $m_\pi^2 = 0$, as necessary for a Goldstone
boson.

It is illuminating to rewrite the couplings in terms of these masses.
The equation of motion can be written as
\begin{equation}
m_{\eta'}^2 - m_\pi^2 = m_{a_0}^2 - m_{\sigma}^2 
\; .
\label{eq:EOM_masses}
\end{equation}
Although here the pion mass vanishes, this relation remains valid even
when $m_\pi^2 \neq 0$, 
Eq. (91) of Ref. \cite{pisarski_chiral_2016}.  Using this relation,
we can express all three parameters in terms of $\phi$ and two masses,
\begin{eqnarray}
m^2 &=& \frac{1}{6} \left( m^2_{\eta'} - 3 \, m_\sigma^2 \right) \; ,
\nonumber \\
\kappa \, \phi &=& \frac{1}{3} \, m_{\eta'}^2  \; ,
\nonumber \\
\lambda \, \phi^2 &=&
\frac{1}{12} 
\left(
m_{\eta'}^2 + 3 \, m_{\sigma}^2 
\right)
\; .
\label{eq:parameters_in_terms_masses}
\end{eqnarray}
As expected, the $\eta'$ is massive because of the determinental coupling
$\sim \kappa$.  Notice that the expectation value $\phi$ is not
fixed by these relations; usually that is determined by the value
of the pion decay constant.  

Usually, in mean field theory one assumes that only the mass parameter
$m^2$ is a function of temperature, and takes $\kappa$
and $\lambda$ to be constant.  
At high temperature 
$m^2(T) \sim \lambda T^2$, but the dependence is more complicated
at small temperature.  We do not need to know this
dependence to determine the masses and couplings
at the chiral phase transition, $T_\chi$.

The solutions to the equation of motion are
\begin{equation}
\phi(T) = \frac{\kappa}{4 \lambda}
\left( 1 \pm \sqrt{1 - \frac{8 \, \lambda}{\kappa^2} \, m^2(T)
} \right)
\; ,
\end{equation}
where in the broken phase the minimum corresponds to the $+$ sign.
The transition occurs when the free energy, which is minus the potential,
is equal to that in the symmetric phase.  
Since ${\cal V}_\Phi(0) = 0$, this occurs when
\begin{equation}
{\cal V}_\Phi(\phi(T_\chi)) = 0 \;\;\; \Rightarrow \;\;\;
\frac{ 8 \, \lambda}{ \kappa^2}
\; m^2(T_\chi) = + \, \frac{1}{9} \; 
\; .
\end{equation}
Just below the transition temperature, 
\begin{equation}
T = T_\chi^-:
\;\;\; 
m_{\eta'} = \sqrt{\frac{\kappa^2}{\lambda}}
\;\;\; , \;\;\;
m_{\sigma} = \frac{1}{3} \; m_{\eta'}
\;\;\; , \;\;\;
m_{a_0} = \sqrt{\frac{10}{9}} \; m_{\eta'}
\; .
\end{equation}
For $T> T_\chi^+$, all masses, including those for the pion, $= m(T)$.  
Assuming that $m^2(T)$ is monotonically increasing with
temperature, in order to have a phase transition 
we need that $m^2(T_\chi) > m^2(0)$.  

The precise mass spectrum for QCD with three massless
flavors is not known at present.  
The relations in Eq. (\ref{eq:parameters_in_terms_masses})
show that we can always treat two masses as free parameters.
As an example, consider 
\begin{equation}
T=0: \; \;
m_{\eta'} = 960
\;\; , \;\;
m_\sigma = 600
\;\; , \;\;
m_{a_0} = 1130
\;\; , \;\;
m^2 = - (162)^2
\; ,
\label{eq:zero_temp_toy}
\end{equation}
where all masses are in MeV.  In this we assume that the $\eta'$
and the $\sigma$ mesons have the values given above, and 
stress that they are only meant as
{\it suggestive}.  The mass of the $a_0$ meson follows from
Eq. (\ref{eq:EOM_masses}), and $m^2$ from 
Eq. (\ref{eq:parameters_in_terms_masses}).
In QCD the masses of the $\eta'$ and the $a_0$ are very close, but
the above value for $m_{a_0}$ isn't so unreasonable in the chiral limit.

Using these values we can then compute the corresponding quantities at
the chiral transition temperature:
\begin{equation}
T=T_\chi: \; \;
m_{\eta'} = 752
\;\; , \;\;
m_\sigma = 251
\;\; , \;\;
m_{a_0} = 835
\;\; , \;\;
m^2 = + (89)^2
\; .
\end{equation}
While all masses decrease with increasing temperature,
at $T_\chi$ those for the $\eta'$ and the $a_0$ 
are still $\sim 75-80\%$ of their values at $T=0$, while
that for the $\sigma$ meson is only $\sim 40\%$.
That doesn't tell us the value of $T_\chi$, since
that depends upon the details of the temperature dependence of $m^2(T)$.

\subsection{Mirror model at zero temperature}
\label{sec:mirror_zero_temp}

We now construct the simplest possible model which illustrates how
two chiral phase transitions arise in the chiral limit.
We assume that the potential for the tetraquark field $\zeta$ is given by
\begin{equation}
{\cal V}_{\zeta}(\zeta)
= m^2 \; {\rm tr} \left( \zeta^\dagger \zeta \right)
+ \kappa \, \left(
{\rm det} \zeta 
+ {\rm c.c.}
\right)
+ \lambda \, {\rm tr} \left(\zeta^\dagger \zeta\right)^2
\; .
\label{eq:toy_zeta}
\end{equation}
We term this the ``mirror'' model, since we choose all parameters
to be identical to those for $\Phi$, Eq. (\ref{eq:toy_Phi}): 
comparing to Eq. (\ref{eq:zeta_pure_three}), we take
$m_\zeta^2 = m^2$, $\kappa_\zeta = \kappa$, and $\lambda_\zeta = \lambda$.

The only term that we include which mixes $\zeta$ and
$\Phi$ is the $Z(3)_A$ invariant term in Eq. (\ref{eq:mix_zeta_phi_three}).
It is not difficult to see that including just this term greatly alters
the mass spectrum.  Taking an expectation value for $\zeta$ which is
diagonal, the mixing term is
\begin{equation}
\langle \zeta^{a b} \rangle = \zeta \, \delta^{a b} 
\;\;\; : \;\;\;
{\cal V}_{{\rm mix}}
= 
3 \,  \widetilde{m}^2 \, \zeta \, \phi 
\; ,
\end{equation}
where again to simplify the notation we
take $\widetilde{m}^2 = m_{\zeta \Phi}^2$.  Henceforth in this section,
by $\zeta$ we denote not the matrix, but the scalar expectation value
of the diagonal component thereof.

From Eq. (\ref{eq:EOM_phi}), the equations of motion become
\begin{eqnarray}
\frac{\partial }{\partial \phi}
\left( {\cal V}_\Phi+ {\cal V}_{{\rm mix}} \right)
&=& 6
\left(
\widetilde{m}^2 \, \zeta
+ \, m^2 \, \phi - \, \kappa \, \phi^2
+ 2 \, \lambda \, \phi^3
\right)
\; ,
\nonumber \\
\frac{\partial }{\partial \zeta}
\left( {\cal V}_\zeta+ {\cal V}_{{\rm mix}} \right)
&=& 6
\left(
\widetilde{m}^2  \, \phi
+ \, m^2 \, \zeta - \kappa \, \zeta^2
+ 2 \, \lambda \, \zeta^3
\right)
\; .
\label{eq:EOM_mirror}
\end{eqnarray}
For each field, the mixing term acts like a background field
proportional to the other field: in the equation of motion for $\phi$,
there is a term $\sim \widetilde{m}^2 \zeta$, and {\it vice versa}.

We first compute the mass spectrum at zero temperature.  At $T=0$,
where $m_\zeta^2 = m_\Phi^2$
the expectation values $\zeta$ and $\phi$ are equal.  The
exact value is not of relevance for our purposes.

The $\Phi$ fields contains two octets, which we term the $\pi$ and $a_0$,
and two singlets, the $\eta '$ and the $\sigma$.
There are similar fields for the $\zeta$,
which we denote as the $\widetilde{\pi}$, $\widetilde{a}_0$, 
$\widetilde{\eta}\,'$, and $\widetilde{\sigma}$.  
The mixing between these fields which is
induced by Eq. (\ref{eq:mix_zeta_phi_three}) is particularly
simple: the 
$\pi$ mixes only with the $\widetilde{\pi}$, the $\eta'$ only with the 
$\widetilde{\eta}\, '$, and so on.  Finding the mass eigenstates
then requires diagonalizing four $2 \times 2$ matrices.
For the pions, using the
equation of motion in Eq. (\ref{eq:EOM_mirror}), 
when $\zeta = \phi$ the mass matrix
between the $\pi$ and the $\widetilde{\pi}$ is
\begin{equation}
{\cal M}^2_{\pi \widetilde{\pi}} = 
\widetilde{m}^2 \; 
\left(
\begin{array}{cc}
-1 & 1 \\
1  & -1 \\
\end{array}
\right)
\; .
\label{eq:mixing_pions}
\end{equation}
The eigenvalues of this matrix are
\begin{equation}
\pi \;  , \; \widetilde{\pi} :
\;\;\; 0
\;\;\; , \;\;\;
 - \, 2 \, \widetilde{m}^2
\; .
\label{eq:masses_mix_pion}
\end{equation}
For the mass squared of the massive ``pion'' to be positive requires that
$\widetilde{m}^2$ is negative.  This is unremarkable, as the
mass squared for both the $\zeta$ and $\Phi$ are also negative.  
Since the expectation values of $\zeta$ and $\phi$ are equal,
after diagonalization each mass eigenstate
is a linear combination of the original fields, in equal proportion.

Since we are in the chiral limit, there is one massless and one massive
octet.  There are nine Goldstone bosons when
$SU(3)_L \times SU(3)_R \times U(1)_A$ symmetry breaks to $SU(3)$.
The quantum breaking of $U(1)_A$
to $Z(3)_A$ makes the $\eta'$ massive and reduces this to eight.
When the $\zeta$ and $\Phi$ are decoupled each has eight Goldstone bosons.
Coupling them makes one of the octets massive, leaving one 
massless octet required by Goldstone's theorem.

The mass squared for the remaining fields are
\begin{eqnarray}
\eta' \; , \; \widetilde{\eta}\, '&:&  \;\;\; 3 \, \kappa \, \phi
\;\;\; , \;\;\;
 3 \, \kappa \, \phi - \, 2 \, \widetilde{m}^2 \; ;
\nonumber \\
a_0 \; , \; \widetilde{a}_0&:&
\;\;\; m^2 + \kappa \, \phi + 6 \, \lambda \, \phi^2  \pm \widetilde{m}^2 \; ;
\nonumber \\
\sigma \; , \; \widetilde{\sigma}&:&
\;\;\; m^2 - 2 \, \kappa \, \phi + 6 \, \lambda \, \phi^2  
\pm \widetilde{m}^2
\; .
\label{eq:masses_mix_mirror}
\end{eqnarray}
These are naturally related to the masses in the absence of mixing,
given in Eq. (\ref{eq:masses_sigma_toy}).  Notice, however,
that the expectation value of $\phi$ is the solution of Eq. 
(\ref{eq:EOM_mirror}), and not the solution of Eq. (\ref{eq:EOM_phi}).

The masses of the $a_0$ and $\widetilde{a}_0$,
and the $\sigma$ and $\widetilde{\sigma}$, are elementary,
just the mass splitting induced 
by off-diagonal elements $\sim \widetilde{m}^2$ in the mass matrix.
The $\eta'$ and $\widetilde{\eta}\,'$ differ from these because
they are Goldstone bosons when $\kappa = 0$.
For all of these mesons, the mass eigenstates are
linear superpositions of the original fields, although not equally.
The exact mixing is easy to work out.

The masses in Eq. (\ref{eq:masses_mix_mirror})
obey the relation
\begin{equation}
m^2_{\eta'} + m^2_{\widetilde{\eta}'} - m^2_\pi - m^2_{\widetilde{\pi}}
= m^2_{a_0} + m^2_{\widetilde{a}_0}
- m^2_\sigma - m^2_{\widetilde{\sigma}} \; .
\label{eq:mirror_chiral_identity}
\end{equation}
One can show that 
given the potentials of Eqs. (\ref{eq:toy_Phi}) and (\ref{eq:toy_zeta}),
this relation remains valid even if we do not assume that the
parameters are related as a mirror model.

We have not checked that this relation remains valid for arbitrary potentials,
but even so it illustrates a more general point.  
The relation for two fields in Eq. (\ref{eq:mirror_chiral_identity})
is very similar to that for one field in 
Eq. (\ref{eq:EOM_masses}).  
For one field, however, there is a puzzle.  As expected, the anomaly
term $\sim \kappa \, {\rm det} \Phi$ 
splits the singlet $\eta'$ from octet $\pi$, making the $\eta'$ heavy.
However, it also pushes the mass of the singlet
$\sigma$ {\it down} relative to the octet $a_0$.  
Now of course to compare to QCD we need to include the effect of quark
masses, especially that the strange quark is heavier than the up and
down.  Even so, it is peculiar
that the {\it singlet} $\sigma$ is lighter than the octet $a_0$ for
the $J^P = 0^+$ field.

With two fields, however, there is no problem, as the only relation is
between the {\it sum} of the masses squared.  Thus the anomaly pushes the 
sum of the mass squared of the $\eta'$
and the $\widetilde{\eta}\,'$ up relative to that for $\pi$ and the
$\widetilde{\pi}$.  Conversely, the anomaly
pushes the masses of both the $\sigma$ {\it and}
the $\widetilde{\sigma}$ down relative to the $a_0$ and
the $\widetilde{a}_0$.  Since the states from the $\zeta$ can be
significantly heavier than the usual states,
it is much easier satisfying this constraint.  See also our 
discussion in Sec. (\ref{sec:mass_tetra}).

\subsection{Mirror model at nonzero temperature}
\label{sec:mirror_temp}

All of these masses in the previous section can only be valid
at zero temperature, where by fiat we imposed the condition that
$m_\zeta^2(0) = m_\Phi^2(0)$.  At nonzero temperature, because the states
are linear combinations of the original fields, with 
mixing due to $\widetilde{m}^2$, then
\begin{equation}
m^2_\zeta(T) \neq m^2_\Phi(T) 
\label{eq:masses_diff}
\end{equation}
at {\it any} nonzero temperature.

This is obvious from effective models.  If one computes
the thermal fluctuations from the $\zeta$ and $\Phi$ fields, 
then just because
the masses of the two multiplets differ, so will the effective
masses for $\zeta$ and $\Phi$.  In other words, we can tune the masses
to be equal at a given temperature, such as zero, but we cannot
impose this naturally at another temperature.

For example, in the limit of high temperature, if we neglect
mesonic fluctuations then the dominant contribution to the thermal
masses are given by quark loops.  For the $\Phi$ field this is
$\sim y_\Phi T^2$, where $y_\Phi$ is the Yukawa coupling of the quarks
to the $\Phi$, Eq. (\ref{eq:chiral_inv_lag}), while for $\zeta$
it is $\sim y_{\zeta 3} T^2$, Eq. (\ref{eq:quark_zeta_three}).
There is no symmetry which relates the two Yukawa couplings, and
so $y_\Phi \neq y_{\zeta 3}$.  Indeed, since the coupling 
$\sim y_{\zeta 3}$ respects the axial $Z(3)_A$ symmetry but not $U(1)_A$,
$y_{\zeta 3}$ vanishes as $T\rightarrow \infty$, while $y_\Phi$ is nonzero.
Thus the two masses 
differ as $T \rightarrow \infty$, Eq. (\ref{eq:masses_diff}).

As an example, we assume that we fix the expectation values
at zero temperature to agree with the value of the pion decay constant
in QCD, which is $\phi(0) = 93/2$, Eq. (93) of Ref.
\cite{pisarski_chiral_2016}.  
From Eqs. (\ref{eq:parameters_in_terms_masses}) and
(\ref{eq:zero_temp_toy}),
\begin{equation}
\phi(0)  = \zeta(0) =46.
\;\;\; , \;\;\;
\kappa = 6680.
\;\;\; , \;\;\;
\lambda = 79.
\; .
\end{equation}
These values are similar to those from a fit to QCD,
Eqs. (95) and (96) of Ref. \cite{pisarski_chiral_2016}.  
Notice that in a linear sigma model
that the couplings $\kappa$ and $\lambda$ are so large because
the pion decay constant is much smaller than the masses of the $\eta'$
and the $\sigma$.

\begin{figure}
\includegraphics[width=0.6\linewidth]{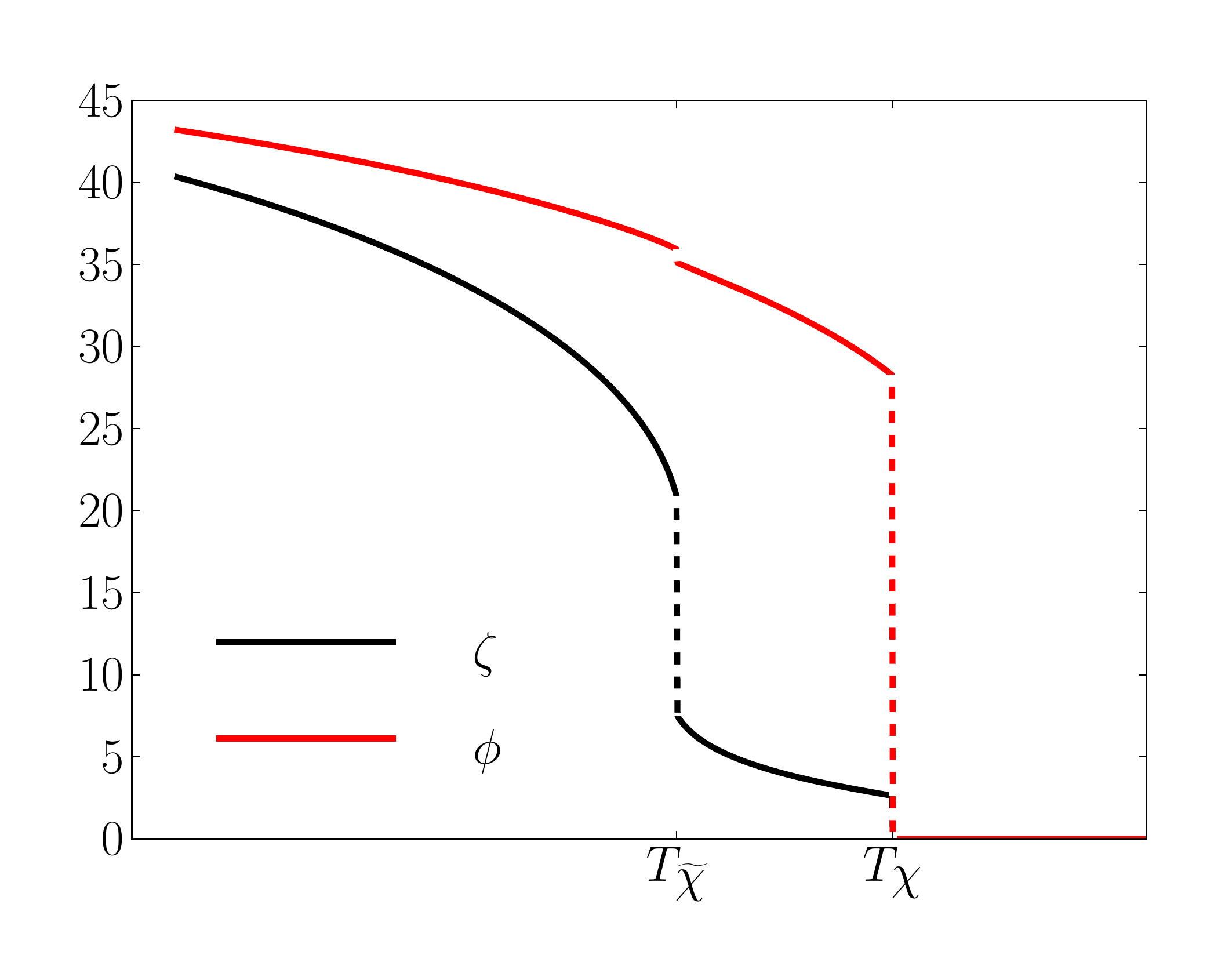}
\caption{
The temperature dependence of the order parameters, $\zeta$ and $\phi$,
in the mirror model for the parameters of
Eq. (\ref{eq:mirror_case_one}).  
There are two chiral transitions of first order, at $T_{\widetilde{\chi}}$ and
then $T_\chi$.
}
\label{fig:mirror_case_one}
\end{figure}

We consider three cases to illustrate the range of possibilities.
In the first case, we take
\begin{equation}
m_\phi^2(T) = 3 \, T^2 \, + m^2(0) 
\;\;\; , \;\;\;
m_\zeta^2(T) = 5 \, T^2 \, + m^2(0) 
\;\;\; , \;\;\;
m_{\zeta \Phi}^2 = - (100)^2 
\; .
\label{eq:mirror_case_one}
\end{equation}
We stress that the temperature dependence is meant only to be illustrative.
At low temperatures massless pions give a contribution $\sim T^2$,
but the other contributions from massive fields are Boltzmann.  
In this instance, the order parameters behave as in 
Fig. (\ref{fig:mirror_case_one}).
There are two first order phase transitions: at $T_{\widetilde{\chi}}$, both
$\zeta$ and $\phi$ jump from one nonzero value to another.  At $T_\chi$,
both jump from nonzero values to zero.  

\begin{figure}
\includegraphics[width=0.6\linewidth]{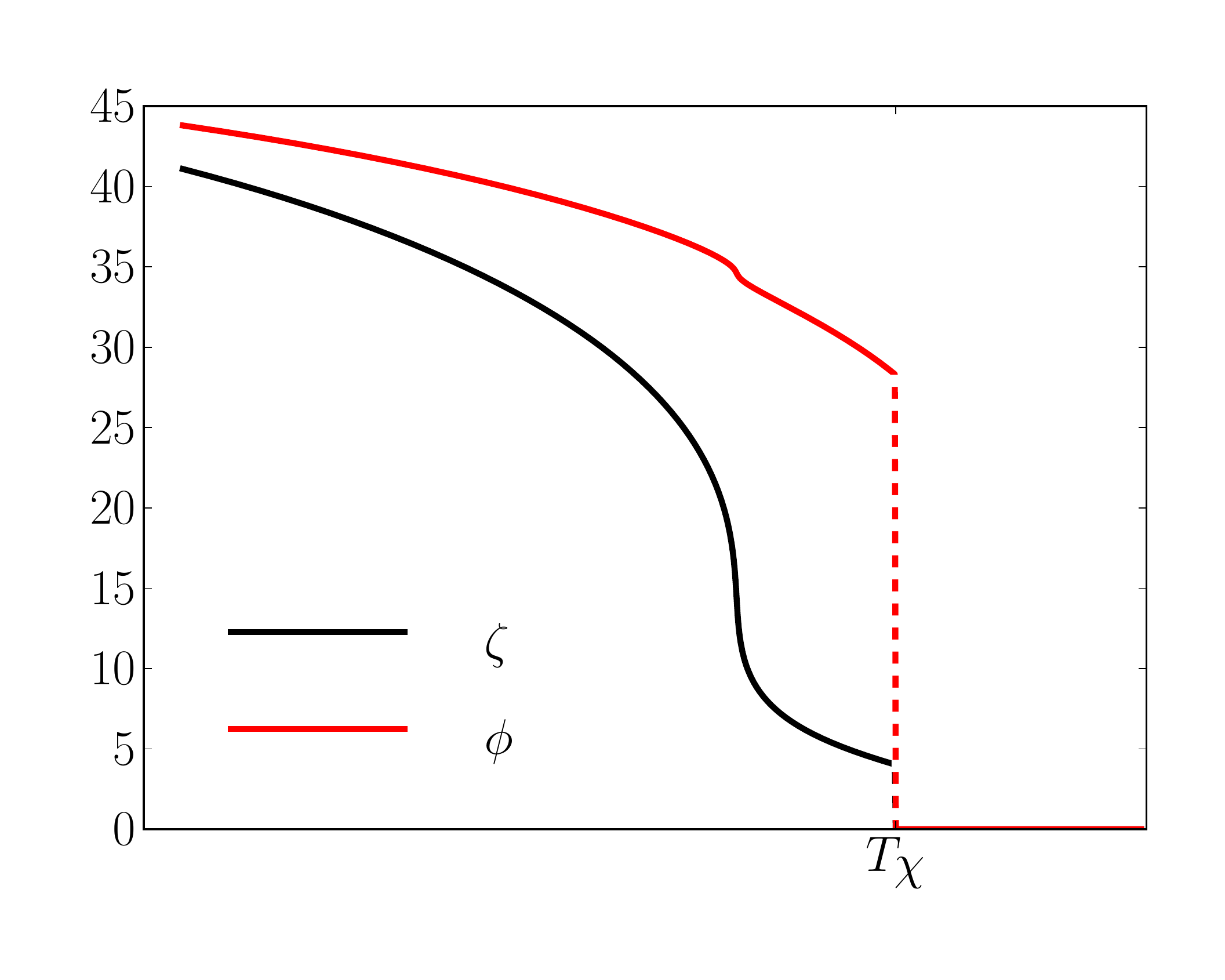}
\caption{
The temperature dependence of the order parameters, $\zeta$ and $\phi$,
in the mirror model, with the parameters of
Eq. (\ref{eq:mirror_case_two}).  
There is one chiral phase transition of first order at $T_\chi$.
}
\label{fig:mirror_case_two}
\end{figure}

In the second case, we only change the mixing mass $m^2_{\zeta \Phi}$,
\begin{equation}
m_\phi^2(T) = 3 \, T^2 \, + m^2(0) 
\;\;\; , \;\;\;
m_\zeta^2(T) = 5 \, T^2 \, + m^2(0) 
\;\;\; , \;\;\;
m_{\zeta \Phi}^2 = - (120)^2 
\; .
\label{eq:mirror_case_two}
\end{equation}
The order parameters behave as in Fig. (\ref{fig:mirror_case_two}).
Because the mixing mass $m_{\zeta \Phi}^2$ is larger, 
the mixing term acts as a larger background field.  This smooths out
the would be transition at $T_{\widetilde{\chi}}$ 
from first order to crossover.
There is then a single chiral phase transition at $T_\chi$.

These examples are only meant to illustrate what is possible,
and should only be taken as such.
Nevertheless, it clearly is possible to obtain a second chiral
phase transition from the presence of the tetraquark condensate.

So far we have only considered the chiral limit.  To understand the
broader implications for the phase diagram of QCD we need to 
consider how nonzero quark masses affect a tetraquark condensate
and the phase diagram.  

\section{Massive quarks}
\label{sec:massive}

\subsection{Mass terms for tetraquarks}
\label{sec:mass_tetra}

To describe QCD it is necessary to include terms for
the explicit breaking of chiral symmetry.  
Let the current quark masses be
\begin{equation}
{\cal M} = {\rm diag} (m_u \, , \, m_d \, , \, m_s ) 
\; ,
\label{eq:mass_phi}
\end{equation}
where $m_u$, $m_d$, and $m_s$ are the masses for the up, down, and strange
quarks.   

Since $m_u$ and $m_d$ are much less than other scales in QCD,
we take the isospin symmetric limit with $m_u = m_d$.  In a sigma
model, the breaking of chiral symmetry is represented including
a background field proportional to the mass matrix,
\begin{equation}
{\cal V}^1_{\Phi}
= - {\rm tr} \left( H_\Phi \left( \Phi^\dagger + \Phi \right) \right)
\; ,
\end{equation}
with 
\begin{equation}
H_\Phi = \left( h_u \, , \, h_u \, , \, h_s \right)
\; .
\label{eq:hphi}
\end{equation}
If chiral symmetry is approximately valid we expect that
the ratio of the $h$'s
is proportional to that for the current quark masses, $h_u/h_s = m_u/m_s$.  
However, the overall constant is given by the details of the fit
to the sigma model.  

For small quark masses, it suffices to include only terms linear in
$H$ and $\Phi$.  A complete catalog of all possible terms is given in
Appendix A of
Refs. \cite{fariborz_toy_2005} and \cite{fariborz_model_2007}.

To understand the leading mass term for the tetraquark field,
imagine computing it explicitly in perturbation theory.
This is of course a terrible approximation, but it should suffice
to get the leading powers of the quark mass right.
For the usual $\Phi$ field, its expectation value
is proportional to a quark loop, 
$\sim {\rm tr} (1/(\slash \!\!\!\! D + m_{quark}))$.
For small masses this trace is proportional to $m_{quark}$, 
so $\langle \Phi^{i i} \rangle \sim m_{quark}$, which is
given by taking $H^{i i} \sim m_{quark}$, Eq. (\ref{eq:hphi}).

The tetraquark field, however, involves the antisymmetric tensor for flavor,
Eqs. (\ref{eq:diquark_twoflavor}) and 
(\ref{eq:diquark_threeflavor}).  For example, the expectation value
of the strange-strange component of the tetraquark field, $\zeta^{s s}$,
involves the product of an up quark loop times a down quark loop.
For small quark masses each is proportional to the mass,
so $\langle \zeta^{s s}\rangle \sim m_u \, m_d$.
The other components follow similarly.
Hence for the tetraquark field, the leading
term which breaks the chiral symmetry is proportional to the
{\it square} of the quark masses, and is
\begin{equation}
{\cal M}_\zeta = \; {\rm diag} \left( m_d \, m_s \; , \; m_u \, m_s 
\; , \; m_u \, m_d \right)
\; .
\end{equation}
Assuming $SU(2)$ isospin symmetry, 
\begin{equation}
{\cal M}_\zeta \approx m_u 
\; {\rm diag} \left( m_s \; , \; m_s \; , \; m_u \right)
\; .
\label{eq:mass_tetra}
\end{equation}
Thus to the linear sigma model we add
\begin{equation}
{\cal V}^1_{\zeta}
= - {\rm tr} \left( H_\zeta \left( \zeta^\dagger + \zeta \right) \right)
\; .
\end{equation}
Assuming $SU(2)$ isospin symmetry,
\begin{equation}
H_\zeta = h \left( h_s \, , \, h_s \, , \, h_u \right)
\; .
\label{eq:back_field_zeta}
\end{equation}
There is no reason for the background field for $\zeta$ to be identical
to that for $\Phi$, and so while we expect that
in $H_\zeta$ we have $h \sim h_u$, we should take $h$ as an
independent constant to be fit by hadronic phenomenology.

If we can neglect mixing, then the mass term of Eq. (\ref{eq:back_field_zeta})
immediately gives us insight into why tetraquarks are so appealing
in QCD.  As we discussed at the end of Sec. (\ref{sec:mirror_zero_temp})
following Eq. (\ref{eq:EOM_mirror}), a sigma model with a single field
$\Phi$ gives a light $\sigma$ which has a large strange component.
For the tetraquark field, however, the mass term 
for the strange-strange component of $\zeta$ is proportional to
the product of the {\it light} quark masses, 
$\sim m_u \, m_d$.  That is, a mass term such as 
Eq. (\ref{eq:mass_tetra}) naturally gives an ``inverted'' mass ordering
which appears to be present in QCD for the lightest $0^+$ multiplet
\cite{jaffe_multiquark_1977, *jaffe_multiquark_1977-1, *jaffe_color_2000,
*jaffe_diquarks_2003, *jaffe_exotica_2005, black_mechanism_2000, fariborz_toy_2005, fariborz_two_2008, fariborz_note_2008, black_putative_1999, *fariborz_model_2007, *fariborz_global_2009, *fariborz_probing_2011, *fariborz_chiral_2011, *fariborz_chiral_2014, *fariborz_probing_2015, tornqvist_lightest_2002, *close_scalar_2002, *amsler_mesons_2004, *napsuciale_chiral_2004, *napsuciale_chiral_2004-3, *maiani_new_2004, *pelaez_light_2004, *pennington_can_2007, *t_hooft_theory_2008, *pelaez_chiral_2011,*heupel_tetraquark_2012, *mukherjee_low-lying_2012, *chen_1_2015, *eichmann_light_2016, giacosa_mixing_2007, parganlija_vacuum_2010, *giacosa_spontaneous_2010, *janowski_glueball_2011, *parganlija_meson_2013, *wolkanowski_scalar-isovector_2014, *janowski_is_2014, *ghalenovi_masses_2015, *giacosa_mesons_2016, olive_review_2014-1, *pelaez_controversy_2015}.

This of course neglects mixing between the $\Phi$ and $\zeta$ fields,
in particular through the direct mixing in Eq. 
(\ref{eq:mix_zeta_phi_three}).  This term does induce an expectation
value $\langle \zeta \rangle \sim m^2_{\zeta \Phi} \langle \Phi \rangle
\sim m^2_{\zeta \Phi} H_\Phi$.  
As we stressed in Sec. (\ref{sec:mirror_zero_temp}), however,
{\it all} fields are linear combinations of $\Phi$ and $\zeta$.
Different choices for the parameters of the model gives different
ratios of mixtures.  

It is still meaningful to stress that the leading term in quark
masses for the tetraquark field is that of Eq. (\ref{eq:back_field_zeta}).
For example, in the limit of high temperature the mixing
term $\sim m^2_{\zeta \Phi}$ is very small, and the breaking of
the chiral symmetry from explicit quark masses is much smaller for
$\zeta$ than for $\Phi$.  

\subsection{Phase diagram for three light flavors}

In this section we make discuss the implications for the phase diagram
in moving away from the chiral limit.  As seen in the discussion
of the mirror model in Sec. (\ref{sec:mirror_temp}), in the chiral
limit it is possible to obtain two chiral phase transitions, at
$T_{\widetilde{\chi}}$ and $T_\chi$.  

A useful way of plotting the phase diagram versus the quark masses
is in the two dimensional plane of the light quark
mass, taking $m_u = m_d$, versus the strange quark mass $m_s$.

When all quark masses are large, there is a region of
first order phase transitions which are
dominated by that for deconfinement.
In a matrix model \cite{kashiwa_critical_2012} the critical line
which borders this region of first order deconfining phase
transitions is determined by the 
color $Z(3)$ field generated by heavy quarks.  Whatever bound
states the heavy quarks form --- whether of two, four, or 
however many quarks ---
seems unlikely to affect
the position of the critical line for deconfinement.

\begin{figure}
\includegraphics[width=0.4\linewidth]{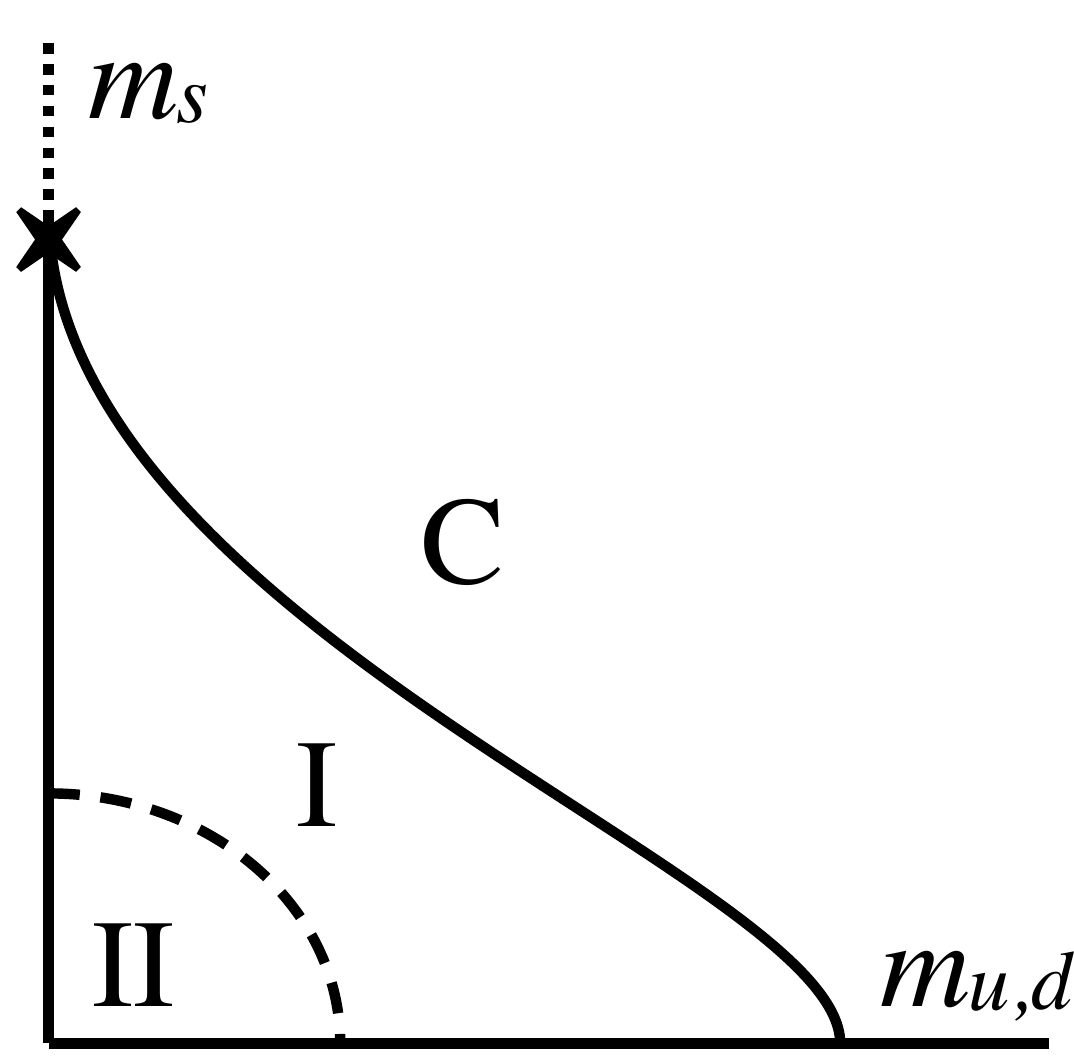}
\caption{
The phase diagram for three light flavors of quarks, in the plane of
$m_u = m_d$ versus $m_s$.  In region II there are two
first order chiral phase transitions; in region I,
one transition; in region C, there is only crossover.
There are critical lines separating regions II and I, and I and C.
QCD lies in the crossover region.  
}
\label{fig:columbia}
\end{figure}

Thus we concentrate on the region of small quark masses.
{\it If} there are two chiral phase transitions in the chiral
limit, $m_u = m_d = m_s$, then it is natural that this persists for
a nonzero width in the plane of $m_u$ and $m_s$.  We illustrate this
in the ``Columbia'' phase diagram of 
Fig. (\ref{fig:columbia}).  
Thus region II denotes where there are two chiral phase transitions
for first order, ending in the dotted line.  In region I, there is
one chiral transition of first order, ending in the solid line.
QCD lies in C, the crossover region.

Both the dotted and solid lines are regions where there is a critical
line.  That there is a critical line in going from two to one chiral
phase transition can be guessed from the behavior of the order parameters
in Fig. (\ref{fig:mirror_case_two}).  Along the dotted critical line,
at a temperature $T_{\widetilde{\chi}}$ there is a linear combination
of the $\Phi$ and $\zeta$ fields which is critical.  This transition
is separate from the first order chiral transition at $T_\chi$.  

The most interesting part of Fig. (\ref{fig:columbia}) is the
left most axis, where 
\begin{equation}
m_u = m_d = 0 \;\;\; , \; \; \; m_s \neq 0
\;\;\; \Rightarrow \;\;\;
H_\Phi = (0,0,h_s) 
\; \;\; , \; \; \;
H_\zeta = (0,0,0)
\; .
\label{eq:left_columbia}
\end{equation}
Notice that $H_\zeta$ vanishes because it is proportional to $m_u$,
Eq. (\ref{eq:back_field_zeta}).  

Consider first the strange-strange component of the tetraquark
field, $\zeta^{s s}$.  Because of mixing with $\Phi$, it develops a
nonzero expectation value.  Then $\zeta^{s s}$ acts
exactly like the tetraquark field for two flavors.  For
instance, the $U(1)_A$ invariant trilinear coupling 
$\sim \kappa_\infty$ for three flavors,
Eq. (\ref{trilinear_u1_sym_three}), reduces directly to the
$U(1)_A$ invariant trilinear coupling for two flavors $\sim \kappa_\infty$
in Eq. (\ref{eq:phi_zeta_twofl_u1}).  In agreement with our
arguments about two flavors in Sec. (\ref{sec:tetra_two}), we do
not expect that the strange-strange component of the tetraquark field
significantly affects the chiral transition when $m_u = m_d = 0$,
Eq. (\ref{eq:left_columbia}).

When $m_u = m_d = 0$ there is one subtlety 
which is worth noting.  Assume that the effects
of the anomaly are large, so that we can assume that there is only
a $Z(2)_A$ symmetry, and not $U(1)_A$.  
For the quark masses as in Eq. (\ref{eq:left_columbia}), from
the $SU(3)_L \times SU(3)_R$ fields $\zeta$ and $\Phi$ we can
obviously extract two $O(4)$ fields, $\vec{\zeta}$ and $\vec{\phi}$.
The effective Lagrangian for these two $O(4)$ fields is 
\begin{eqnarray}
{\cal V}_{m_u = m_d = 0}
&=& 
\; m^2_\phi \; \vec{\phi}^{\, 2} + 
m^2_\zeta \; \vec{\zeta}^{\,2} + m^2_{\zeta \phi} 
\; \vec{\zeta} \cdot \vec{\phi}
\nonumber \\
&+& \lambda_\phi \, ( \vec{\phi}^{\, 2} )^2
+ \lambda_\zeta \, ( \vec{\zeta}^{\, 2} )^2
+ \lambda_{\zeta \phi 1}\, \vec{\zeta}^{\, 2} \, \vec{\phi}^{\, 2} 
+ \lambda_{\zeta \phi 2} \, ( \vec{\zeta} \cdot \vec{\phi} )^2
\; .
\label{eq:pot_zero_mu_md}
\end{eqnarray}
The couplings above are obviously related to those we denoted
previously.  Clearly, there is no trilinear coupling between
$\vec{\zeta}$ and $\vec{\phi}$ which is $O(4)$ invariant.

It is not difficult to convince oneself that having two fields
doesn't alter the standard picture.  For small $m_s$, in regions
II and I, there are either two or one chiral phase transition.
Approaching the boundary of region I from below, there is a tricritical
point at $m_s^{tri}$, denoted by a cross in Fig. (\ref{fig:columbia}).
For $m_s > m_s^{tri}$, there is a line of second order phase transitions,
in the universality class of $O(4)$.  
Unless there is a conspiracy in the masses
for $\vec{\zeta}$ and $\vec{\phi}$, though, 
having two fields doesn't alter the critical behavior in the least:
it simply means that some 
linear combination of $\vec{\zeta}$ and
$\vec{\phi}$ is the relevant critical field.
The shape of the critical
line between regions I and C as $m_s \rightarrow 0$ is dictated
by the tricritical behavior.  There is no such curvature for the
critical line, separating regions I and II, because the transition remains
of second order as one moves along the critical line
when $m_s \rightarrow 0$. 

Of course this assumes that there is a region II with two chiral transitions
of first order.  This question can only be settled definitively by
numerical simulations on the lattice.  Since QCD only
finds a crossover, such simulations need to be done for very light
quarks, which is
most challenging.  Nevertheless, if the lattice does find two chiral
phase transitions for light quarks, 
would be strong if indirect evidence for
the effects of tetraquarks in QCD.

\section{Phase diagram in $T$ and $\mu$}
\label{sec:phase_diagram_T_mu}

In QCD, at nonzero temperature but zero quark chemical potential,
numerical simulations on the lattice indicate that there is
no true phase transition, but only a crossover for a single chiral
transition at a temperature of $T_\chi \sim 155$~MeV
\cite{cheng_transition_2006, *bazavov_equation_2009, *cheng_qcd_2008, *fodor_phase_2009, *aoki_qcd_2009, *borsanyi_qcd_2010, *borsanyi_is_2010, *cheng_equation_2010, *bazavov_chiral_2012, *bhattacharya_qcd_2014, *bazavov_equation_2014, *borsanyi_full_2014, buchoff_qcd_2014, ding_thermodynamics_2015, *ratti_lattice_2016}.  
Thus any second chiral transition associated with
the tetraquark field, at $T_{\widetilde{\chi}} < T_\chi$, 
is almost certainly a crossover.  

Even so, at nonzero temperature and quark chemical potential,
there is naturally a relation between the crossover
line for tetraquark field and the transition line for color superconductivity.
A tetraquark field is important because of diquark pairing,
with the most attractive channel for quark-quark scattering being
antisymmetric in both flavor and color,
as a type of generalized Breit interaction 
\cite{jaffe_multiquark_1977, *jaffe_multiquark_1977-1, *jaffe_color_2000,
*jaffe_diquarks_2003, *jaffe_exotica_2005}.
Thus it is hardly suprising that in considering the scattering
of two quarks at the edge of the Fermi sea at nonzero density,
that color superconductivity occurs in the corresponding, most
attractive channel.  

The analogy is deeper.  Consider the diquark operators for two flavors, 
$\chi_L^A$ in Eq. (\ref{eq:diquark_twoflavor}),
and three flavors, $\chi^{a A}_L$ in Eq. (\ref{eq:diquark_threeflavor}).
These are almost identical to the operators which condense
when color superconductivity occurs
\cite{schaefer_continuity_1999, pisarski_superfluidity_1999, *pisarski_why_1999, *pisarski_gaps_2000, pisarski_critical_2000, alford_color_2008}.
That is, the tetraquark field is {\it directly} the gauge invariant
square of the diquark operators, Eqs. 
(\ref{eq:define_zeta_two}) and (\ref{eq:define_zeta_three8}).
Of course the tetraquark field must be a color singlet, 
since it appears in the confined phase at zero
temperature and chemical potential; there is no evidence for a color
superconducting phase in vacuum.

There are differences between the condensation of a tetraquark field
in vacuum and color superconductivity.  Color superconductivity
is dominated by the scattering of quarks at opposite edges of the
Fermi surface, between two quarks with momenta $+\vec{p}_F$ and
$-\vec{p}_F$.  For a tetraquark condensate, the entire tetraquark
field carries zero momentum, but each diquark operators carries equal
and opposite momenta.  Further, the color-flavor locking which occurs
for three flavors and three colors 
\cite{schaefer_continuity_1999, alford_color_2008}
has no analogy for the tetraquark condensate.

Even so, as one moves out in quark chemical potential, then it is
reasonable to speculate that a crossover line for the tetraquark
condensate connects smoothly with that for color superconductivity.  

\begin{figure}
\includegraphics[width=0.5\linewidth]{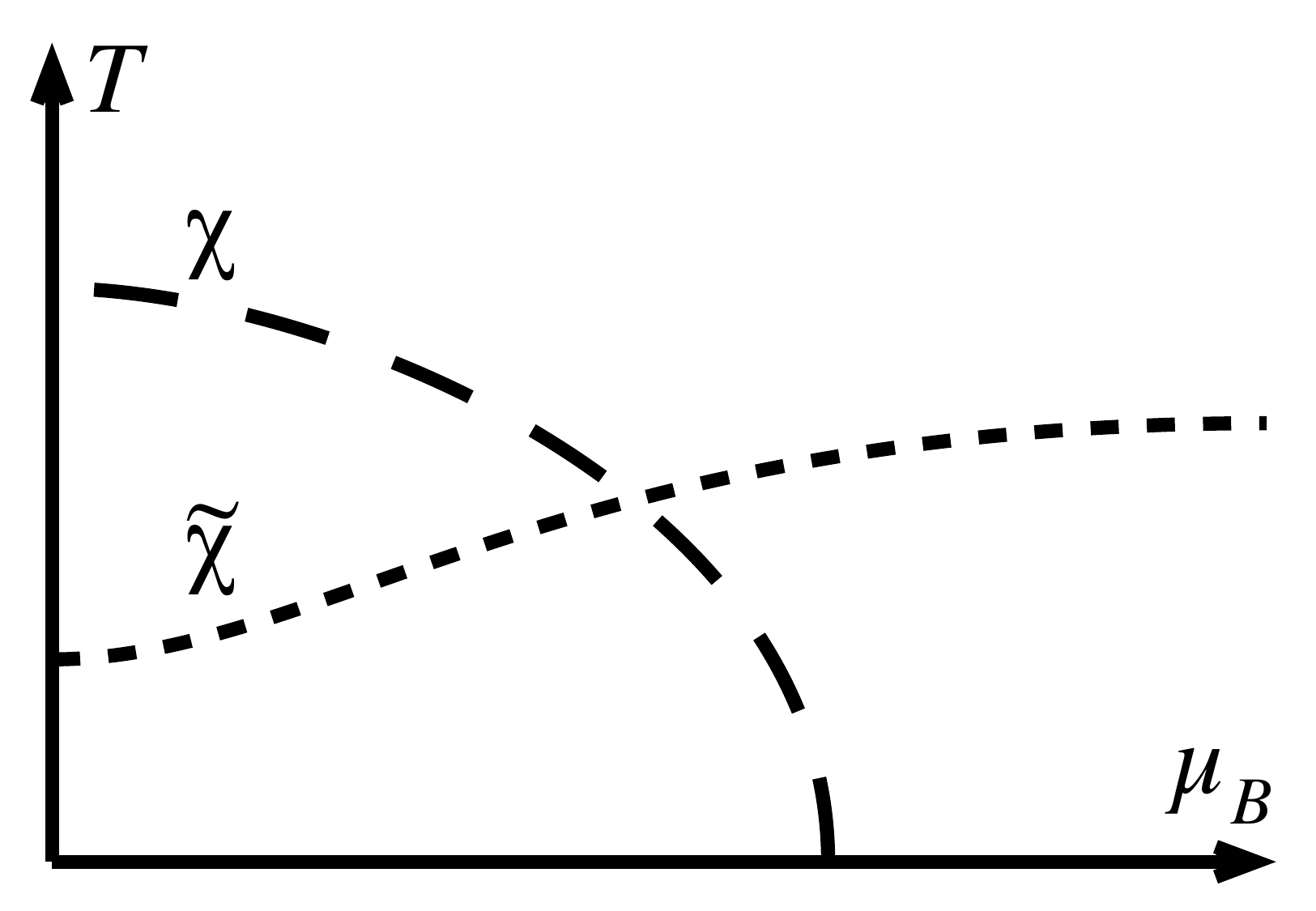}
\caption{
A conjectured phase diagram in temperature $T$ and baryon 
chemical potential $\mu_B$. The $\chi$ line is that for the
chiral transition.  The $\widetilde{\chi}$ line is for a second
transition related to the presence of the tetraquark condensate,
which may connect smoothly to the transition line for
color superconductivity.  
}
\label{fig:T_mu}
\end{figure}

We illustrate this in Fig. (\ref{fig:T_mu}), as a cartoon
of the possible phase diagram.  In particular, we do not indicate
whether the transitions are crossover, or true phase transitions,
of either first or second order.  The chiral crossover line at $\mu_B = 0$
may end in a critical endpoint, and then turn first order
\cite{stephanov_signatures_1998, *stephanov_event-by-event_1999}.  The 
transition line for color superconductivity probably includes a segment
which is a line of second order phase transitions 
\cite{pisarski_critical_2000}.
Further, it is not evident how the tetraquark/color
superconducting line is related to that for hadronic superfluidity
through a confined but dense quarkyonic phase \cite{mclerran_phases_2007}.

\section{Four flavors}
\label{sec:four_flavors}

We conclude with elementary comment about operators for four flavors
and three colors.  The diquark operator which directly generalizes
those for two and three flavors is
\begin{equation}
\chi^{(a b) A}_L =
\epsilon^{a b c d} \; \epsilon^{A B C} \;
( q^{c B}_L )^T  \; {\cal C}^{-1} \; q^{d C}_L \;  .
\label{eq:diquark_fourflavor}
\end{equation}
This operator is an antisymmetric two-index tensor in the $SU(4)_L$ flavor
group, a $\overline{\bf 6}$.  As before we then
combine a left handed diquark field with
a right handed diquark to form a color singlet tetraquark field,
\begin{equation}
\zeta^{(a b),(c d)}
= 
\left(\chi^{(a b) A}_R\right)^\dagger \; \chi^{(c d) A}_L
\; ,
\end{equation}
where $\zeta^{(a b),(c d)} = - \zeta^{(b a ),(c d)}$, {\it etc.}, 
so $\zeta$ lies in the 
$\overline{{\bf 6}} \times {\bf 6}$ representation of 
$SU(4)_L \times SU(4)_R$.  

Since the tetraquark field $\zeta$ 
lies in a higher representation of the chiral
symmetry group than $\Phi$, there is no direct mixing between
them.  There is a cubic coupling,
\begin{equation}
i \, 
\left( \Phi^{a a'} \right)^* \; 
\left( 
\zeta^{(a b),(a' b')} 
- \left(\zeta^{(a b),(a' b')}\right)^* \; 
\right)
\; \Phi^{b b'}
\; ,
\label{eq:cubic_four}
\end{equation}
where the overall factor of $i$ follows from the antisymmetry
of $\zeta$.  This coupling is invariant under $Z(4)_A$ but not $U(1)_A$.
There are many quartic couplings which can be written down, including
$U(1)_A$ invariant terms such as
\begin{equation}
{\rm tr} \left( \Phi^\dagger \Phi \right)
{\rm tr} \left( \zeta^\dagger \zeta \right)
\label{eq:quartic_fourA}
\; .
\end{equation}
There are also quartic couplings which are
invariant only under $Z(4)_A$, such as 
\begin{equation}
\Phi^{a a'} \; \zeta^{(a b),(a' b')} \; \zeta^{(b c),(b' c')} \;
\Phi^{c c'} + {\rm c.c.}
\label{eq:quartic_fourB}
\; .
\end{equation}
Consequently, when $\Phi$ develops a vacuum expectation value it
affects $\zeta$.  Even so, there are no couplings present which would
indicate that the presence of $\zeta$ materially affects either the
pattern of symmetry breaking, or its restoration, in any significant
way.

This is a more general phenomenon.  Starting with QCD, from the
quark fields it is possible to construct a ladder of operators
related to chiral symmetry breaking.  The simplest is $\Phi$, 
which transforms as the fundamental representation in $SU(N_f)_L \times
SU(N_f)_R$, where $N_f$ is the number of flavors.  In addition,
there are four quark operators, six quark operators, and so on.  
Catagorizing them according to
the representations of the chiral symmetry group, typically they are either
singlets, such as ${\rm tr}( \Phi^\dagger \Phi)$, or 
transform according to higher representations.  
Singlet fields are like the tetraquark for two flavors in 
Sec. (\ref{sec:tetra_two}), which don't dramatically affect things.
Similarly, fields in higher
representations do couple to fields in lower representations,
but again it is unnatural for them to have any dramatic effect.
This is just because fields in high representations have more indices, and so
as illustrated
in Eqs. (\ref{eq:cubic_four}),
(\ref{eq:quartic_fourA}), and (\ref{eq:quartic_fourB}),
need more $\Phi$'s to absorb all of them.

The one exception is if there is another field
which transforms in
the fundamental representation which directly mixes with $\Phi$.
If one abandons the prejudice of only
using diquarks fields, then it is possible to construct such
a field for four flavors and three colors.
For a large number of colors, Rossi and Veneziano suggested
that junctions, which couple all colors together through
an antisymmetric tensor in color, matter
\cite{rossi_string-junction_2016}.  
This suggests using junctions in both color and flavor to
form a triquark operator
\begin{equation}
\chi^{a}_L =
\epsilon^{a b c d} \; \epsilon^{A B C} \;
q^{b A}_L \; ( q^{c B}_L )^T  \; {\cal C}^{-1} \; q^{d C}_L \;  .
\label{eq:triquark_fourflavor}
\end{equation}
This is a color singlet, and since it is composed of
three quark fields, transforms as a fermion.  

We can naturally
combine a left handed triquark with a right handed triquark
to form a hexaquark state
\begin{equation}
\xi^{a b} =
\left(\chi^a_R\right)^\dagger \; \chi^b_L
\; ,
\end{equation}
which transforms as $\overline{{\bf 4}} \times {\bf 4}$
under $SU(4)_L \times SU(4)_R$.  

The analysis of coupling the hexaquark field $\xi$ to the
usual chiral field $\Phi$ is very similar to that for three flavors.
The axial $U(1)_A$ symmetry is reduced to $Z(4)_A$ by the anomaly,
with $\Phi$ carrying axial charge $= +1$, and
$\xi$, axial charge $=-3$.

The $U(1)_A$ invariant couplings include mass and quartic
couplings between $\xi$ and $\Phi$, in direct analogy to
Eqs. (\ref{eq:phi_pure_three}), (\ref{eq:zeta_pure_three}), and
(\ref{quartic_u1_sym_threefl}), replacing $\zeta \rightarrow \xi$.
For three flavors there is the $U(1)_A$ invariant determinental term
of Eq. (\ref{trilinear_u1_sym_three}).  
The analogous term for four flavors is 
\begin{equation}
\kappa_\infty
\; 
\epsilon^{a b c d} \;
\epsilon^{a' b' c' d'}
\; \xi^{a a'} \; \Phi^{b b'} \; \Phi^{c c'} \; \Phi^{d d'}
+ {\rm c.c.}
\; .
\end{equation}

Of the couplings which are invariant under $Z(4)_A$ but not $U(1)_A$,
the most important is a direct mixing term
\begin{equation}
m^2_{\xi \Phi}
\; {\rm tr} 
\left(
\, \xi^\dagger \, \Phi + \, \Phi^\dagger \, \xi
\right)
\; ,
\end{equation}
as in Eq. (\ref{eq:mix_zeta_phi_three}).  The determinental terms
$\sim \kappa_\Phi \, {\rm det} \Phi$ and 
$\sim \kappa_\xi \, {\rm det} \xi$ are of quartic order for
four flavors.  There are two determinental terms of quartic order,
\begin{equation}
\epsilon^{a b c d} \;
\epsilon^{a' b' c' d'}
\left(
\kappa_{\xi \Phi 1}
\; \xi^{a a'} \; \xi^{b b'} \; \Phi^{c c'} \; \Phi^{d d'}
+ \kappa_{\xi \Phi 2}
\; \xi^{a a'} \; \xi^{b b'} \; \xi^{c c'} \; \Phi^{d d'}
\right) + {\rm c.c.}
\; .
\end{equation}
The other quartic terms invariant under $Z(4)_A$ are 
those of Eq. (\ref{eq:quartic_zeta_phi_z3}), just replacing 
$\zeta \rightarrow \xi$.

The analysis of the chiral transition for four flavors with a hexaquark field
is then closely analogous to that with a tetraquark field for three flavors.
The hexaquark field $\xi$ mixes directly
with $\Phi$, so if one field condenses both do.
Similarly, the restoration of chiral symmetry involves both fields.
For four flavors
the determinental terms are of quartic instead of cubic order, and so
do not automatically generate first order transitions.

However, it is also possible that chiral transitions are driven
first order by fluctuations, where coupling constants flow
from positive to negative values.  
This occurs to leading order in an $\epsilon$-expansion about $4-\epsilon$
dimensions when $N_f > \sqrt{2}$ \cite{pisarski_remarks_1984}.
It is not clear if this remains true in three dimensions: for
two flavors, there is evidence that a new critical point develops
for $SU(2)_L \times SU(2)_R \times U(1)_A = O(4) \times O(2)$
\cite{pelissetto_relevance_2013, *nakayama_approaching_2014, *nakayama_bootstrapping_2015, meggiolaro_remarks_2013, *fejos_fluctuation_2014, *fejos_functional_2015, *sato_linking_2015, *eser_functional_2015}.
If so, it is possible that such a new critical point persists up to four
flavors.  We note that even with two fields in the fundamental representation,
only one linear combination of the two contributes to the putative
critical behavior at $T_\chi$.  If the transition is of first
order, it is also possible that there are having two chiral fields
in the fundamental representation produces two chiral transitions,
as for three flavors.

It is interesting to speculate what the relevant effective fields are
for the chiral transition when the number of colors, $N_c$, is
greater than three.  
It has been suggested that tetraquarks persist in the usual
large $N_c$ limit, where $N_f$ is held fixed as 
$N_c \rightarrow \infty$
\cite{weinberg_tetraquark_2013, *knecht_narrow_2013, *lebed_large-n_2013, *cohen_quantum-number_2014, *cohen_non-ordinary_2014, *maiani_tetraquarks_2016}.
On the other hand, our analysis suggests that 
the relevant limit might be more general: taking
$N_f = N_c \rightarrow \infty$, instead of tetraquarks
we would use junctions in both flavor and
color to form $2(N_c - 1)$ quark states which transform in the fundamental
representation of the chiral symmetry group.

\section{acknowledgements}
We thank D. Rischke and J. Schaffner-Bielich for 
emphasizing to us that tetraquarks might be relevant to
the chiral phase transition.
R.D.P. thanks the U.S. Department of Energy for support
under contract DE-SC0012704.

\bibliography{qks}
\newpage

U.S. Department of Energy Office of Nuclear Physics or High Energy Physics

{\it Notice:} 
This manuscript has been co-authored by employees of Brookhaven 
Science Associates, LLC under Contract No. DE-SC0012704 with 
the U.S. Department of Energy. The publisher by accepting the manuscript for 
publication acknowledges that the United States Government retains a 
non-exclusive, paid-up, irrevocable, world-wide license to publish or 
reproduce the published form of this manuscript, or allow others to do so, 
for United States Government purposes.
This preprint is intended for publication in a journal or proceedings.  
Since changes may be made before publication, it may not be cited or 
reproduced without the author’s permission.

{\it DISCLAIMER}:
This report was prepared as an account of work sponsored by an agency of the 
United States Government.  Neither the United States Government nor any 
agency thereof, nor any of their employees, nor any of their contractors, 
subcontractors, or their employees, makes any warranty, express or implied, 
or assumes any legal liability or responsibility for the accuracy, 
completeness, or any third party’s use or the results of such use of any 
information, apparatus, product, or process disclosed, or represents that 
its use would not infringe privately owned rights. Reference herein to any 
specific commercial product, process, or service by trade name, trademark, 
manufacturer, or otherwise, does not necessarily constitute or imply its 
endorsement, recommendation, or favoring by the United States Government or 
any agency thereof or its contractors or subcontractors.  The views and 
opinions of authors expressed herein do not necessarily state or reflect 
those of the United States Government or any agency thereof. 

\end{document}